\documentclass[bibyear]{aa}
\usepackage{graphicx}
\begin{document}
   \title{The long bar as seen by the VVV Survey: II. Star counts}

   \author{E. B. Am\^{o}res
          \inst{1,2,3}
          \and
          M. L\'{o}pez-Corredoira\inst{4,5}
          \and
          C. Gonz\'{a}lez-Fern\'{a}ndez\inst{6}
          \and
           A. Moitinho\inst{3}
          \and
          D. Minniti\inst{7,8,9}
          \and
          S. Gurovich \inst{10,11}
          }
        \institute{UEFS, Departamento de F\'{i}sica, Av. Transnordestina, S/N, Novo Horizonte, Feira de Santana, BA, Brazil, CEP
         44036-900 \\ \email{ebamores@uefs.br}
         \and
         UEFS, Observat\'{o}rio Astron\^{o}mico Antares, Rua da Barra, 925, Jardim Cruzeiro, Feira de Santana, BA, Brazil, CEP 44024-432 \\
         \and
        SIM, Faculdade de Ci\^encias da Universidade de Lisboa, Ed. C8. Campo Grande 1749-016 Lisbon,
        Portugal \\
         \and
             IAC, V\'{i}a L\'{a}ctea s/n, E38200 - La Laguna (Tenerife),  Spain\\
         \and
             Departamento de Astrof\'{i}sica, Universidad de La Laguna, E-38206,
             La Laguna, Tenerife, Spain \\
         \and
             Departamento de F\'{i}sica, Ingenier\'{i}a de Sistemas y Teor\'{i}a de la Se\~{n}al,
             Universidad de Alicante, Apdo. 99, E-03080 Alicante, Spain
         \and
             Departamento Astronom\'{i}a y Astrof\'{i}sica, Pontificia Universidad
             Cat\'{o}lica de Chile, Av. Vicu\~{n}a Mackenna 4860, Santiago, Chile\\
         \and
             Vatican Observatory, Vatican City State V-00120, Italy. \\
         \and
             Departamento de Ciencias F\'{i}sicas, Universidad Andres Bello,
             Santiago, Chile \\
         \and
             Observatorio Astron\'{o}mico de C\'{o}rdoba, Universidad Nacional de
             C\'{o}rdoba, Laprida 854, 5000 C\'{o}rdoba, Argentina \\
          \and
             Instituto de Astronom\'{i}a Te\'{o}rica y Experimental
            (IATE-CONICET), Laprida 922 X5000BGR C\'{o}rdoba, Argentina \\
             }
   \date{Received 20 June 2012 / accepted 20 June 2013}

  \abstract
   {There is still some debate about the presence and the morphological properties of the long
   bar in the inner Galaxy.}
   {We investigate the morphological properties of the long Galactic bar using the VVV
   survey extending star counts at least 3 mag deeper than 2MASS. Our study covers the relatively unexplored negative longitudes of the Galactic
   bar. We obtain a detailed description of the spatial distribution of star counts towards the long Galactic bar as well as to measure
   its parameters.}
   {We performed star counts towards $-20^\circ < \ell < 0$, $|b| \leq 2^\circ$ using VVV,
   2MASS, and GLIMPSE data. We applied an average interstellar extinction correction.
   We also adjusted latitudinal profiles to obtain the centroid variation and bar thickness.}
   {We probe the structure of long Galactic bar, as well as its far edge at $\ell \approx -14^\circ $.
   The differences between counts with and without extinction correction allow us to produce a crude extinction map showing regions with
   high extinction, mainly beyond the end of long Galactic bar. The latitudinal profiles show evidence of the centroid vertical variation with
   Galactic longitude reaching a minimum at $\ell \approx -13.8^\circ$. The bar has an inclination angle $\alpha =43^\circ \pm 5^\circ$ with respect to the line
   Sun-Galactic center. In addition, we have determined the bar parameters, such as thickness, length, and stellar distribution.}
   {}

   \maketitle
%

\section{Introduction}

The understanding of the structure that can be attributed to a bar
in the Galactic center has significantly advanced in the three
past decades, mainly thanks to the IRAS, COBE/DIRBE, TMGS, DENIS,
2MASS, and UKIDSS  infrared surveys. These surveys allowed us to
increase the knowledge obtained from the Hayakawa et al. (1981)
and Matsumoto et al. (1982) maps, not only of the structure, but
also of the shape of the Galactic center (Blitz \& Spergel 1991,
hereafter BL91).

After studying the asymmetries in the Galactic center, Lizst \&
Burton (1980) and Burton \& Lizst (1983) suggested they could be
attributed to a bar with an inclination angle of approximately
$25^\circ$. BL91 argued for the existence of asymmetries in the
photometric images of the Galactic bulge by analyzing the 2.4 $\mu
$m observations performed by Matsumoto et al. (1982) and of
Weiland et al. (1994) based on the images obtained by the
COBE/DIRBE experiment in 1.25, 2.2, 3.5, and 4.9 $\mu $m. The
asymmetry is also visible in star counts (e.g., Stanek et al.
1994, Hammersley et al. 1994), which show systematically more
stars at positive Galactic longitudes (within $<30^\circ$) and
close to the Galactic plane, compared to negative longitudes. It
was also supported by the distribution of red clump stars
(Hammersley et al. 2000, Cabrera-Lavers et al. 2007a), the
dynamics of the stellar and the gaseous components in the Galactic
center (e.g., Minchev et al. 2007), gravitational micro-lensing
toward the Galactic bulge (Paczynski et al. 1994 and Popowski et
al. 2005), and the kinematic effects on the local disk stars
(Gardner \& Flynn 2010; Romero-G\'omez et al. 2011).

While some papers refer to a thick triaxial structure (bulge or
thick bar) with a semi-major axis length of 2.5 kpc and a position
angle (PA) of 15$^\circ $--30$^\circ $ with respect to the
Sun-Galactic center direction (e.g., L\'opez-Corredoira et al.
2005; Habing et al. 2006; Rattenbury et al. 2007; Vanhollebeke et
al. 2009; Robin et al. 2012), other contributions suggest that
there is also a long thin bar, the in-plane bar, with a half
length of 4 kpc and a position angle of around 45$^\circ$: e.g.,
L\'opez-Corredoira et al. (2007, LC07 and references therein),
Cabrera-Lavers et al. (2007a), Vallenari et al. (2008), Churchwell
et al. (2009).

This long bar has a tip in the positive longitude at the beginning
of the Scutum's arm (Dame et al. 1986). As realized by Sevenster
et al. (1999), the angle is around $25^\circ$ or lower when low
latitudes are excluded from the fit, and around 40$-$45$^\circ$ in
the plane regions within $-15^\circ < \ell < 30^\circ$.
Furthermore, there is some research on a possible third component,
a nuclear bar (Alard 2001, Nishiyama et al. 2005, Gonz\'alez et
al. 2011) within $|\ell| < 4^\circ $ in the plane. Some of us
think that this is a more uncertain structure, since it is very
dependent on a correct subtraction of the bulge+long bar, the
excess of very bright stars due to star forming regions in the
inner bulge (L\'opez-Corredoira et al. 2001b), or the errors
affecting the distance of the red clump stars (see Appendix in
LC01).

Rather than two large structures, it was suggested by
Mart\'inez-Valpuesta \& Gerhard (2011) a scenario of a single
structure with a twisted major axis. In principle, we find this
proposal acceptable from a purely morphological point of view. In
any case, the concepts are different: bars or pseudo-bulges form
as a result of instability in differentially rotating disks
(Sellwood 1981), whereas bulges are primordial Galactic
components. Therefore, the scenario bulge+bar is not the same
thing as a single structure, and one should observe other
differences apart from the morphology. We think that the
particular features of this new proposal of a \emph{single twisted
bulge/bar} scenario leaves certain observational facts
unexplained, such as the star formation regions at the tips of the
long bar, whereas the model of a misaligned bulge + long bar
successfully explains them (L\'opez-Corredoira et al. 2011).

In addition, there was evidence of kinematic differences between
metal-rich and metal-poor populations in the bulge (Minniti et al.
1996). We now know that there are two distinct populations in the
thick bulge: a metal-rich population with bar-like kinematics and
a metal-poor population with kinematics corresponding to an old
spheroid or a thick disk, so the two main scenarios for the bulge
formation co-exist within the Milky Way bulge (Babusiaux et al.
2010).

The largest (deepest) near-infrared survey of the Milky Way (MW)
disk and bulge is the VISTA Variables in the Via L\'actea,
hereafter VVV (Minniti et al. 2010, Saito et al. 2012), an
important observational base for studying the Galactic structure.
Works with VVV data, such as those of Saito et al. (2011) and
Gonz\'alez et al. (2011, 2012), have already provided new insight
in to the inner structure of our Galaxy.

Here we use the combined 2MASS, VVV, and GLIMPSE products to
explore the long bar at negative longitudes ($-20^\circ < \ell
<0$, $|b| \leq 2^\circ$), which has been less explored than the
positive longitude counterpart. In particular, star counts are
performed in the $K_{\rm  s}$ band for investigating the shape and
structure of the bar at negative longitudes and computing its
parameters (axis dimension, edges, size, inclination angle, etc.).
Our study follows the approach of L\'opez-Corredoira et al.
(2001a, hereafter LC01), who performed a detailed analysis of star
counts in the long bar region ($|\ell|\le 30^\circ $, $|b| \leq
2^\circ $) based on TMGS and DENIS data. A comparison with the
different extinction methods used for general extinction
corrections (Majewski et al. 2011 and LC01) is also presented.

This is the second in a series of two papers addressing the
Galactic bar as seen by the VVV survey. Here, we focus on results
from star counts. The first paper is dedicated to the analysis of
color-magnitude diagrams (Gonz\'alez$-$Fern\'andez et al. 2012,
hereafter Paper I).

The remainder of this paper is organized as follows. Section 2
presents the VVV, 2MASS, and GLIMPSE data used in this work,
including calibration, source matching in both surveys,
elimination of duplicated stars in the VVV tiles, and elaboration
of the VVV catalogs. Section 3 presents the method for correcting
the effects of interstellar extinction in 2MASS, VVV, and GLIMPSE
data and for defining their completeness limits. An analysis of
the latitudinal and longitudinal\footnote{latitudinal profiles are
the distribution of star counts vs. latitude and longitudinal
profiles are the distribution of star counts vs. longitude.} star
counts profiles is presented in Section 4. Section 5 is dedicated
to the resulting bar maps, parameters, and their discussion.
Finally, Section 6 addresses the conclusions of this study and
gives some final remarks.


\section{Data}

\subsection{VISTA Variables in the Via L\'actea (VVV)}

VVV is an ESO public variability survey with the 4-m VISTA
telescope at Cerro Paranal (Minniti et al., 2010). It performs
observations in the Z, Y, J, H, and $K_{\rm  s}$ near infrared
bands towards the Galactic bulge and part of the disk, covering a
total area of 562 square degrees. VVV is in its fourth year. Full
disk and bulge coverage in Z, Y, J, H, and $K_{\rm s}$ bands has
been secured. For a detailed description of the observations
during the first year of the survey, see Saito et al. (2012). The
data were reduced using the CASU
pipeline\footnote{http://casu.ast.cam.ac.uk/}.

\begin{figure*}[t]
\centerline{
\includegraphics[width=11cm,angle=90]{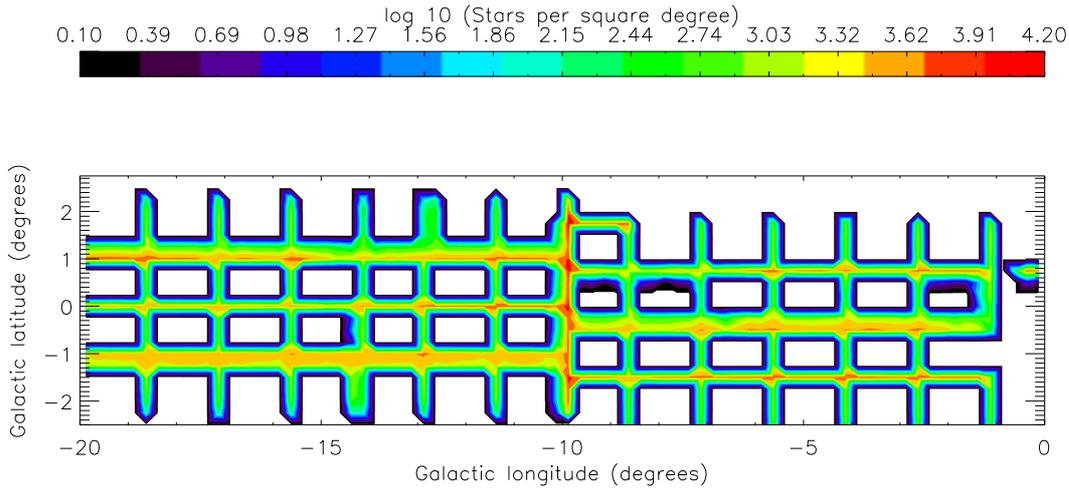}}
\caption[]{\label{figure1}{Tiles with overlap regions. Star counts
binned in ($\Delta l=\Delta b=0.25^\circ$). Colors represent the
number of duplicated stars in log. White means no duplicated
stellar counts in that area}.}
\end{figure*}

This work makes use of 56 tiles covering a region of approximately
90 square degrees. The tile identification labels and their
coordinates are presented in Saito et al. (2012, Table A.1). It is
well established that long bar ends near $\ell \sim -15^\circ$
(LC01 and references therein). For this reason, our analysis is
restricted to Galactic longitudes within $\ell \sim -20.0^\circ$
to 0.0$^\circ$.

We take the VVV tile catalogs toward the bar (including disk and
bulge) in each individual J, H, and $K_{\rm  s}$ filters (version
1.1) to create multi band merged catalogs for each tile region. A
band-merging python code was developed that emulates the method
being implemented by VISTA Science Archive with search radius
equal to 1.0 arcsec. For details on the VVV and 2MASS (Skrutskie
et al. 2006) cross-match, see Paper I.

The sources used in this work are all gathered in a single file.
Only sources classified as stellar (VVV-icls classification flag
equal to $-1$) by VVV pipeline in both J and $K_{\rm  s}$ filters
were used. The file includes the source and tile identifiers,
magnitude, magnitude error, and a flag to indicate whether the
source was taken from 2MASS or from VVV.

The VVV and 2MASS photometry of stars in common were compared and
the difference in magnitude for each star in each filter was
calculated. For most tiles, the average differences in J and
$K_{\rm  s}$ were less than 0.1 mag; however, for eight tiles
(b327, b328, b329, b330, b331, b332, d070 and d071), we noticed
differences in K$_s$ ranging from 0.2 to 0.7 mag, both positive
and negative. For these tiles we opted to transform the VVV
photometry to the 2MASS system by adding the corresponding
photometric offsets.

The VVV observational strategy was designed to have observations
of the same sources in a small overlap of contiguous tiles, which
is useful not only for variability studies but also for
calibration purposes. In this way, tiles contain overlap regions
that correspond to 6\% of the observed sources for a given tile.
As we work with star counts, it is necessary to identify these
sources in order to avoid artifacts due to duplication of counts.

Duplicate stars were eliminated as follows: i) for each tile we
computed its limits (lower and higher) for both Galactic longitude
and latitude; ii) from these limits we verified, for each tile,
the possible same sources considering a match of 1 arcsec within a
strip delimited by 0.3 degrees (chosen in order to have a generous
search region for the detection of duplicate stars) for each side
of a rectangle for all other tiles; iii) since one source in other
tile is identified as similar source of the reference tile, we
quoted it as a flag pointing to a duplicated source. Applying this
analysis we have found 1,663,426 sources that were eliminated from
our study leaving 27,687,811 stars with valid photometry for both
J and $K_{\rm  s}$.

Figure 1 shows a map with counts for the overlapping regions. One
can identify that typical counts in the densest regions range from
3,000 to 7,000 stars. A tilt feature can be seen at different
latitude ranges toward $\ell \sim -10^\circ$ that can be
attributed to the fact that tile centers are located at different
latitudes when considering $\ell < -10^\circ$ and $\ell >
-10^\circ$, respectively. White regions mean no duplicated counts.
The counts are irregular, and eliminating duplicated sources is
mandatory for star counts studies toward any large VVV region.

\subsection{GLIMPSE}

The GLIMPSE data (Benjamin et al. 2003 and  Churchwell et al.
2009) for all phases of the project (GLIMPSE I, II and 3D) are
available at IPAC
webpage\footnote{http://irsa.ipac.caltech.edu/data/SPITZER/GLIMPSE}.
As each phase corresponds to specific regions with different
spatial coverage in both longitude and latitude towards the
Galactic plane, we have selected all of them toward $-20^{o} <\ell
<0^{o}$. Once the overlaping regions were identified, we followed
the GLIMPSE team recommendation (see footnote 3) to use the most
recent data, as for instance, GLIMPSE II data for the region that
extends from $\ell \sim -11.0^\circ$ to $-10.0^\circ$ even though
data for this region is available in both GLIMPSE I and II.

Care was taken to limit the overlap regions to avoid duplicate
sources in the star counts analysis. For each phase of the project
there are separated files that correspond to a specific range of
Galactic longitude, one degree wide. We have joined these files in
a single file for each phase with: Galactic coordinates,
magnitudes, and the errors in GLIMPSE filters: [3.6],[4.5],[5.8],
and [8.0] $\mu$m, as well as 2MASS photometry (magnitudes and
errors) in J, H, and $K_{\rm  s}$.

Interstellar extinction is lower in the mid-IR (MIR) bands (e.g.
[8.0] $\mu$m) than in optical or near-IR (NIR). Stellar spectral
energy distributions sampled in the MIR have similar shapes,
independent of stellar type. This allows a more precise
determination of extinction through the use of adequate color
combinations. Based on these ideas, Majewski et al. (2011,
hereafter M11) developed a method termed Rayleigh-Jeans color
excess (RJCE) based on 2MASS and GLIMPSE data providing useful
relations to determine interstellar extinction, see Equation (1)
of M11. Further details on extinction corrections in GLIMPSE data
are given in section 3.2.

Since VVV is deeper than 2MASS, we choose to increase our matched
source catalog by also matching GLIMPSE data against our VVV
source data (combined J, H and $K_{\rm  s}$ photometry).

Because the extinction correction method of M11 also requires the
H magnitude, we have compiled another matched VVV catalog of
stellar sources with reliable photometry in J, or H, or $K_{\rm
s}$ and applied the same procedure as described in the previous
section for identifying duplicate sources in the tiles. This
resulted in a catalog of positions and photometry for 55,793,434
VVV sources (after discarding 6\% of the duplicates).

The difference between the number of sources in this second
catalog in relation to the first one mentioned in Section 2.1 is
the result of it containing all sources with at least one
detection in one filter, contrary to the first one that needs to
have detection for both J and $K_{\rm s}$. It should be noted that
to compute GLIMPSE counts at [8.0] $\mu$m as indicated by Equation
(5), it is mandatory that a source be detected in [4.5], [8.0]
$\mu$m, and H filters. In summary, for the analysis of the
VVV+2MASS (Section 3.1) we only use J and $K_{\rm s}$. However, in
this second catalog only 3\% of the sources are detected for both
J and $K_{\rm s}$ and not at H.

\bigskip
\begin{table*}
\centering \caption{List of the coverage for GLIMPSE data used in
the present work. The columns \textit{2MASS} and
\textit{2MASS$+VVV$} are respectively the number of valid sources
detected in the three filters (J, H and $K_{\rm  s}$);
\textit{only GLIMPSE} means sources detected only by GLIMPSE. For
GLIMPSE II, the latitudes limits expand to b $\pm 1.5$ and $\pm
2.0$ for longitude ranges $-5^\circ \leq \ell \leq -2^\circ$ and
$-2^\circ \leq \ell \leq 0^\circ$, respectively.}
\begin{flushleft}
\begin{tabular}{cccccc}
\hline Data release & Gal. longitude & Gal. latitude & 2MASS & 2MASS$+$VVV & only GLIMPSE\\
\hline \noalign{\smallskip}
GLIMPSE I  & $-20^\circ \leq \ell \leq -11^\circ$ & $|b| \leq 1^\circ$ &  998,451 & 3,253,321 &  4,174,079 \\
GLIMPSE II & $-11^\circ \leq \ell \leq   0^\circ$ & $|b| \leq 1^\circ$ &1,826,965 & 8,320,253 & 11,077,393 \\
GLIMPSE III& $-20^\circ \leq \ell \leq   0^\circ$ & $|b| \leq 2^\circ$ &1,347,374 & 2,999,419 &  4,066,203 \\
\noalign{\smallskip} \hline
\end{tabular}
\end{flushleft}
\end{table*}
\bigskip

Next, we cross-matched this catalog (by selecting the nearest
source within a search radius equal to 1") with the GLIMPSE merged
catalog mentioned above producing another catalog with coordinates
and VVV and GLIMPSE magnitudes (and errors). Table 1 summarizes
GLIMPSE limits for both Galactic longitude and latitude for each
phase of the project used in the present work, as well as the
comparison of the number of sources obtained with common
photometry J, H, and $K_{\rm  s}$ using only 2MASS data as
provided by GLIMPSE merged catalog and VVV+2MASS catalog produced
by us. The number of matches increases significantly using VVV
photometry instead of only GLIMPSE+2MASS.

As can be seen in Table 1, by adding the matched GLIMPSE+VVV+2MASS
source catalog to our analysis, a substantial increase (a factor
$>4$ for GLIMPSE II) is obtained in matched source numbers, as
compared to a 2MASS (only) cross matches. Since use of VVV data
has the advantage of getting to a larger number of fainter and
redder sources, we can expect that the extinction correction will
be significantly more precise.

\section{Extinction and completeness limits}

Interstellar extinction plays an important role in analysis in the
Galactic plane, as shown in Am\^{o}res \& L\'{e}pine (2005,
hereafter AL05) and Robin (2009 and references therein).
Assessment of extinction is essential for a correct star counts
analysis. Without it, analysis can become severely biased,
especially in high extinction areas.

Below, we outline the methods adopted in correcting for extinction
and defining completeness limits for the VVV+2MASS and GLIMPSE
data sets. A detailed analysis of the interstellar extinction
toward VVV fields is beyond the scope of this present paper, and
it is presented in others papers of the VVV collaboration: e.g.,
Gonz\'alez et al. (2012) and Chen et al. (2013).

\subsection{VVV}

We follow the method described in LC01 for determining an
extinction free magnitude. It should be noted that this method is
also quite similar to the one presented by Alard (2002). As
pointed out by LC01, the method used in the present paper is
roughly consistent with that of Schultheis et al. (1999), except
that here each star is individually corrected for extinction
before each area is averaged. Anyway, the method is also validate
considering star counts simulations as presented in Appendix A.

Since the sources with $J-K_{\rm s} < 0.5$ are dominated by local
disk dwarfs (see Robin et al. 2003 and Marshall et al. 2006), we
have not used them in our star counts method. Of course, there
will also be disk stars with $J-K_{\rm s} > 0.5$. However, as
pointed by LC01, this will be a small proportion of the total
sources and their distribution is symmetrical and in principle
predictable, so at most there will be a loss of contrast for the
inner Galaxy features. For the case where the majority of the
sources are relatively concentrated in a certain location along
the line of sight, this is therefore a straightforward method for
recovering the form of the underlying star distribution, at least
in the infrared.

Moreover, this approach is reasonable (LC01) thanks to the high
density of sources in the inner Galaxy. In many regions, over 50\%
of the sources at a particular magnitude are from the inner
Galaxy, so the vast majority of the detected sources come from a
relatively restricted distance range. The correction holds for
stars satisfying $J-K_s \ge (J-K_s)_0$ and is given by

\begin{equation} \label{equ:a8mu}
\displaystyle m_{K_{\rm s}, {\rm correc.}}= m_{K_{\rm
s}}-\frac{A_{K_{\rm s}}}{A_J-A_{K_{\rm s}}}Max[0,(J-K_{\rm
s})-(J-K_{\rm s})_{0}],
\end{equation}
\noindent

\bigskip

\noindent where

\[ \frac{A_{K_{\rm s}}}{A_J-A_{K_{\rm s}}}=\frac{3}{5}.\]

\bigskip

Since some stars will have color bluer than the average of the
whole population, these stars should have non physical negative
extinction. To avoid applying non physical correction, their
extinction values are set to 0, and therefore for stars with
$J-K_{\rm s} <(J-K_{\rm s})_{0}$, the corrected magnitude equals
the observed one: $m_{K_{\rm s}, {\rm correc.}}= m_{K_{\rm s}}$ in
Equation (1).

The value of 0.6 in Equation (1) was obtained considering an
average value from Rieke $\&$ Lebofsky (1985) and Glass (1999)
that obtained 0.64 and 0.55, respectively, for the ratio
$\frac{A_{K_{\rm s}}}{A_J-A_{K_{\rm s}}}$. We have considered
$(A_{V}:A_{J}:A_{K_{\rm s}}=1.0:0.282:0.112)$ and
$(A_{V}:A_{J}:A_{K_{\rm s}}=1.0:0.256:0.089)$ for the Rieke $\&$
Lebofsky (1985) and Glass (1999) extinction coefficients,
respectively.

That we have used a single $(J-K_s)_0$, even considering there is
a transition between the population content of the bar/bulge and
the disk, could be justified by the very slight modification of
the average color. Star formation regions are more abundant in the
disk that has younger populations and an excess of supergiants and
bright giants, but these very bright stars are still much scarcer
than other fainter stars.

The red clumps have more or less the same intrinsic
color\footnote{As pointed out by Sarajedini et al. (1995), the
luminosity of red clumps has a small dependence for both age and
metallicity.}, and will have different reddening from the disk and
from the bulge, but the counts of red clumps are dominated by the
bulge$/$bar. The simulations in our Appendix A illustrate this.

\begin{figure}
\centering
\includegraphics[width=6.0cm,angle=90]{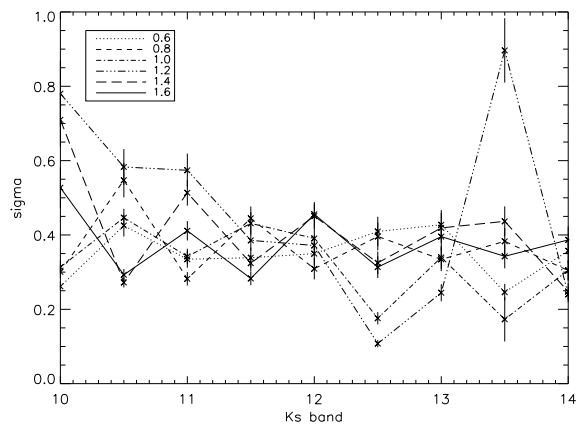}
\caption{$\sigma$ as computed in Equation (2) as a function of the
completeness limit for each value ($J-K_{\rm s})_{0}$ as presented
in the legend for all fields (with area equal to 0.0625 square
degrees) used in the present work.} \label{figure2}
\end{figure}

To verify that adopting $(J-K_{\rm s})_{0} = 1.0$ is a valid value
for the intrinsic color $(J-K_{\rm s})_{0}$ in Equation (1), we
have generated a series of sets of \textit{corrected} $K_{\rm s}$
photometry using $(J-K_{\rm s})_{0}$ ranging from 0.6 to 1.6 at
intervals of 0.2 mag. Each set is split in regions of 0.25$^\circ
\times 0.25^\circ$ square degrees. The area of 0.25$^\circ \times
0.25^\circ$ square degrees for each region was used as a
compromise between having a good spatial resolution (the lower the
better) and having enough counts (the more the better). Moreover,
using fields with this area reduces the effects for both
interstellar extinction and crowding that changes significantly
from neighboring regions in the Galactic plane. Star counts with
0.5 mag steps were then performed for each of these split regions.
The appropriate value for the constant should be the one producing
the lowest variation in the counts between neighboring regions.

Due to field crowding, interstellar extinction and variable
conditions during the observations have an impact on star counts,
so care must be taken to assure that all regions in our analysis
are within their completeness limits. We therefore performed the
analysis considering different magnitude ranges by imposing a
series of magnitude cut-offs ranging from 9.0 to 14.0 (0.5 mag
steps). Then, for each of these cut-offs and for each region, we
computed the squared difference between the counts obtained for
this direction and its eight neighboring regions, as

\bigskip
\begin{equation} \label{equ:rjce}
\displaystyle \sigma _{i,j}^2=\frac{1}{8}\sum _{k,l}^8
\frac{(N_{K,l}-N_{i,j})^2}{N_{i,j}^2}
\end{equation}
\noindent
\bigskip

\noindent where $N(K,l)$ is the number of stars for a given bin,
and $N_{i,j}$ is each neighbor cell in $(\ell,b)$ space.

The final total measure of variation (for each magnitude cut-off)
is defined as the square root of the sums of the square
differences, divided by the number of regions ($n$):

\bigskip
\begin{equation} \label{equ:rjce}
\sigma ^2=\frac{1}{n}\sum _{i,j}^n \sigma _{i,j}^2,
\end{equation}
\noindent
\bigskip

\noindent Figure 2 shows the measure of variation ($\sigma$, from
now on) as a function of the cut-off magnitude for the different
trial values of the correction constant. The value of $(J-K_{\rm
s})_{0} = 1.0$ clearly produces the lowest $\sigma$ up to the
\textit{corrected} $K_{\rm s} \sim 13.5$ mag. It keeps the minimum
for the range 12 $< K_{\rm s} < 14$. The other value $(J-K_{\rm
s})_{0} = 1.2$ is a little bit lower for 12 $< K_{\rm s} < 13$,
but it increases significantly for 13 $< K_{\rm s} < 14$.

The estimate of the completeness limit for each field was
performed using star counts and histograms with bin size equal to
0.5 as a function of magnitude (also referred to as luminosity
function). The peak of each histogram for each field was defined
as the completeness limit. Later, we elaborated a map of the
completeness limit as a function of Galactic latitude and
longitude, which shown that after extinction correction (Equation
1) most of them, e.g., 95 \% of fields have $K_{\rm s} \geq$ 13.5
mag as completeness limit. The fields that have completeness
limits lower than 13.5 mag are mainly located at $|\ell|\le
2^\circ$. Since this region is not the focus of our work, we
preferred to lose completeness for those few fields in order to
have deep completeness along the entire bar.

We also analyzed the counts as a function of magnitude, and the
value of $K_{\rm s} \sim 13.5$ mag represents that the stellar
counts stop growing. In addition, choosing this value as the
completeness limit also provides the best contrast in the star
counts maps of the long bar. Similar analysis was performed in
other NIR filters. The completeness for J and H filters were 17.0
and 16.0 mag, respectively.

\subsection{GLIMPSE}

As pointed out by M11, the use of NIR and MIR photometry has
several advantages since extinction is noticeably lower than in
the visible, and higher precision foreground extinction values can
be determined when combining MIR with NIR photometry. We performed
star counts using the combined VVV+GLIMPSE+2MASS data described in
Section 2.2. As in the previous section, counts were performed
considering grids of 0.25$^\circ \times 0.25^\circ$ square
degrees. Here, extinction is computed with Equation (4). The
completeness limits at [8.0] and [4.5] $\mu$m were determined for
each field (as explained above), which are than 11.0 and 10.5 mag
in most cases. Using the relations provided by Rieke \& Lebofsky
(1985), the extinction at [8.0] $\mu$m is given by

\begin{equation} \label{equ:a8mu}
\displaystyle A([8.0 \mu m]) = 0.020 (A_{K_{\rm  s}}/0.112),
\end{equation}
\noindent

\noindent where $A_{K_{\rm  s}} = 0.112 A_V$.

To better correct the effects of interstellar extinction, we used
Equation (1) provided by M11:

\begin{equation} \label{equ:rjce}
\displaystyle A(K_{\rm  s}) = 0.918 (H-[4.5 \mu m]-0.08)
\end{equation}
\noindent

\noindent The result of the combined VVV+2MASS+GLIMPSE data
dereddened using Equation (4) is presented in Section 5.

\section{Star counts}

The following sections analyze the star counts from two
perspectives: their longitudinal and latitudinal profiles.

\subsection{Longitudinal counts}

One effective way to study counts is to analyze their variation
with longitude (e.g., longitudinal profiles) for different
latitude ranges. In this case we considered three different
ranges: i) $|b| \leq 0.375^\circ$ (b=0$^\circ$, $\pm 0.25^\circ$);
ii) 0.375$^\circ < |b| \leq 1.125^\circ$ (b=$\pm
1.0,0.75,0.50^\circ$); iii) 1.375$^{o}$ $\leq |b| \leq
2.125^\circ$ (b= $\pm 1.5,1.75,2.0^\circ$). Figure 3 (upper left
panel) shows a longitudinal profile for the observed counts
(corrected by interstellar extinction) up to $K_{\rm s}=$ 13.5 mag
for the three latitude ranges mentioned above. The longitude bin
width is 1$^\circ$. A clear decrease in the counts until $\ell
\sim -14^\circ$ for the three latitude ranges can be seen, most
notably for Galactic latitudes located toward $|b| \leq
0.375^\circ$.

\begin{figure*}[t]
\centering
\resizebox{9.0cm}{!}{\includegraphics[angle=90,scale=.90]{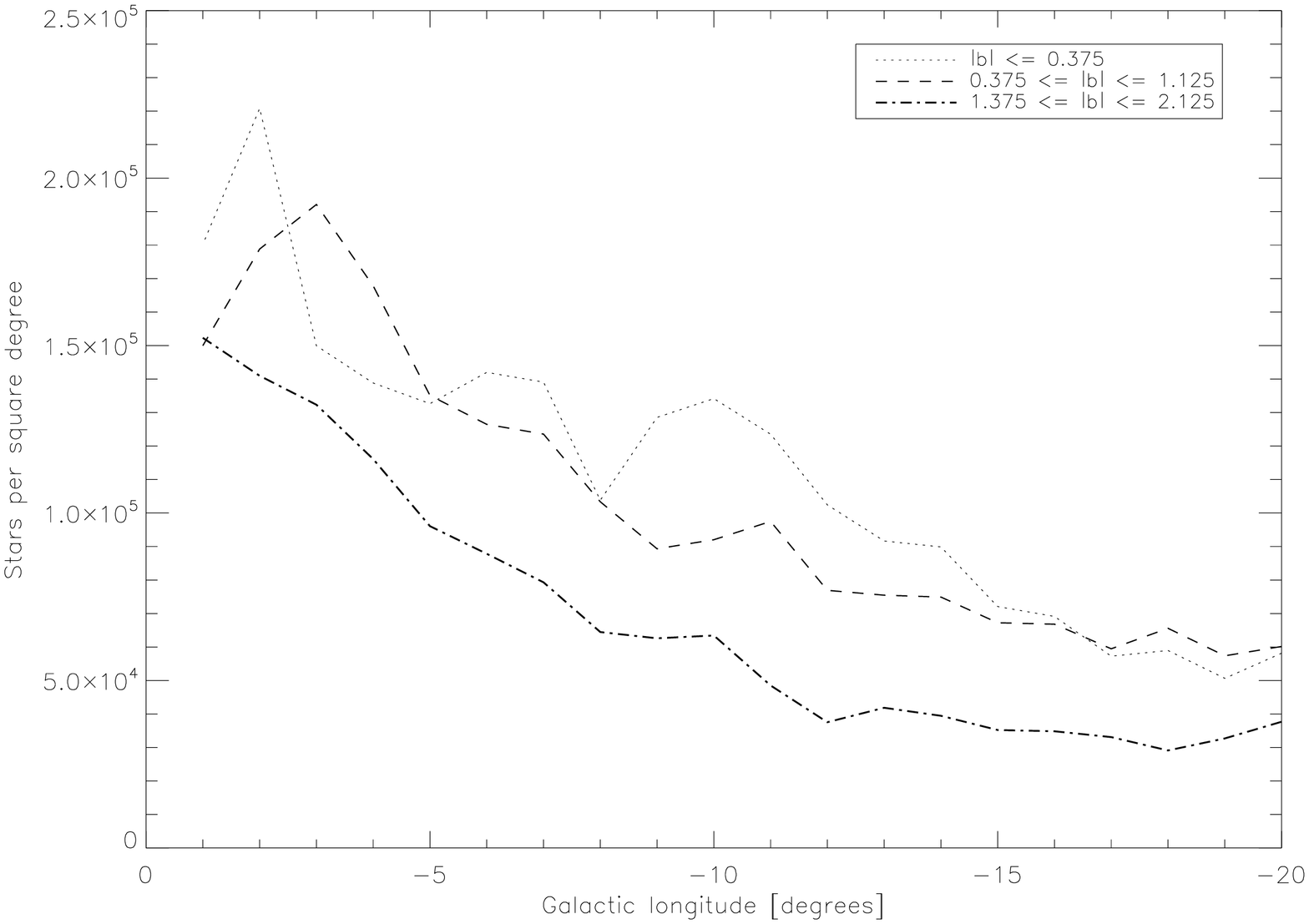}}
\resizebox{9.0cm}{!}{\includegraphics[angle=90,scale=.90]{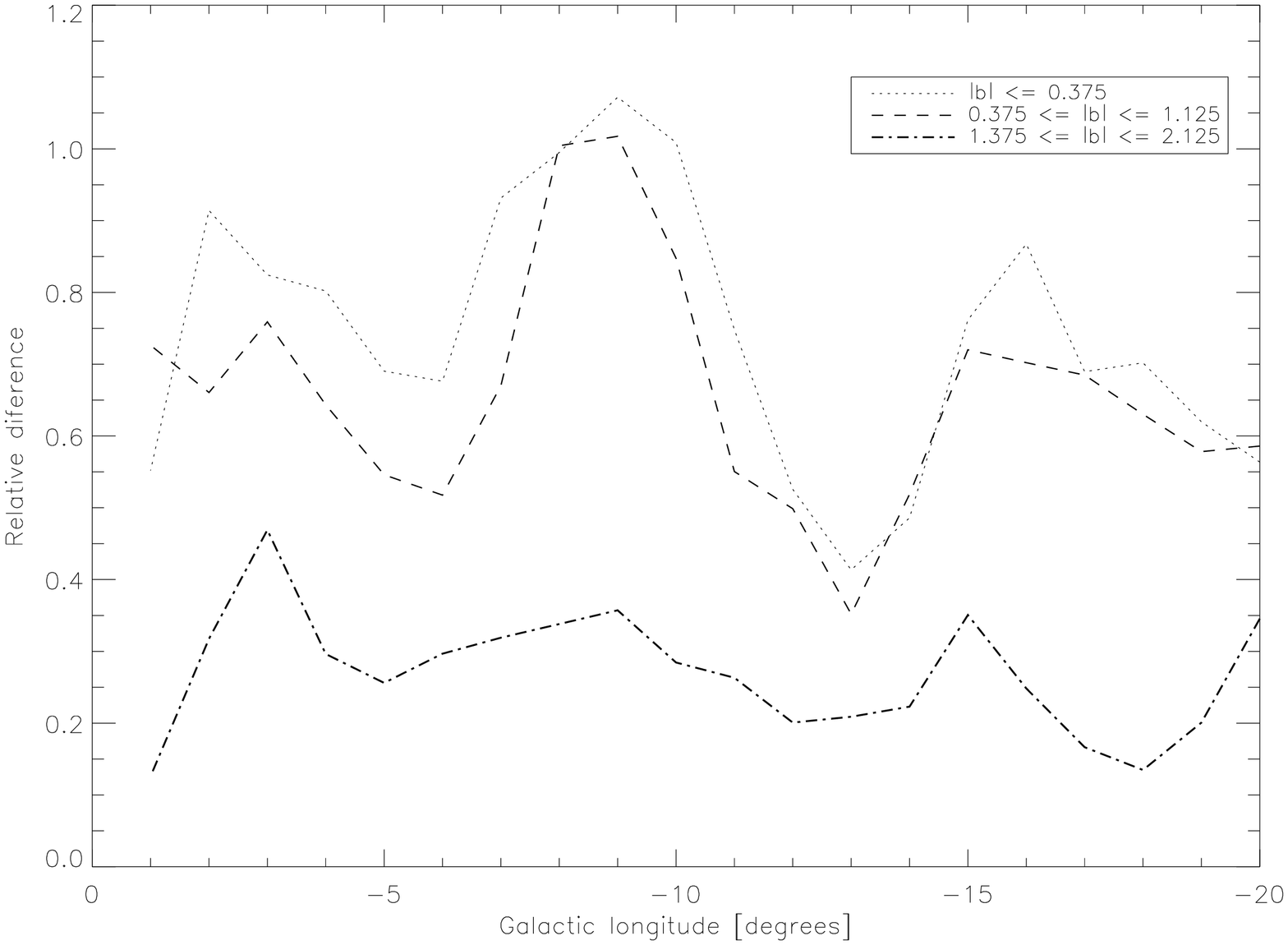}}
\resizebox{9.0cm}{!}{\includegraphics[angle=90,scale=.90]{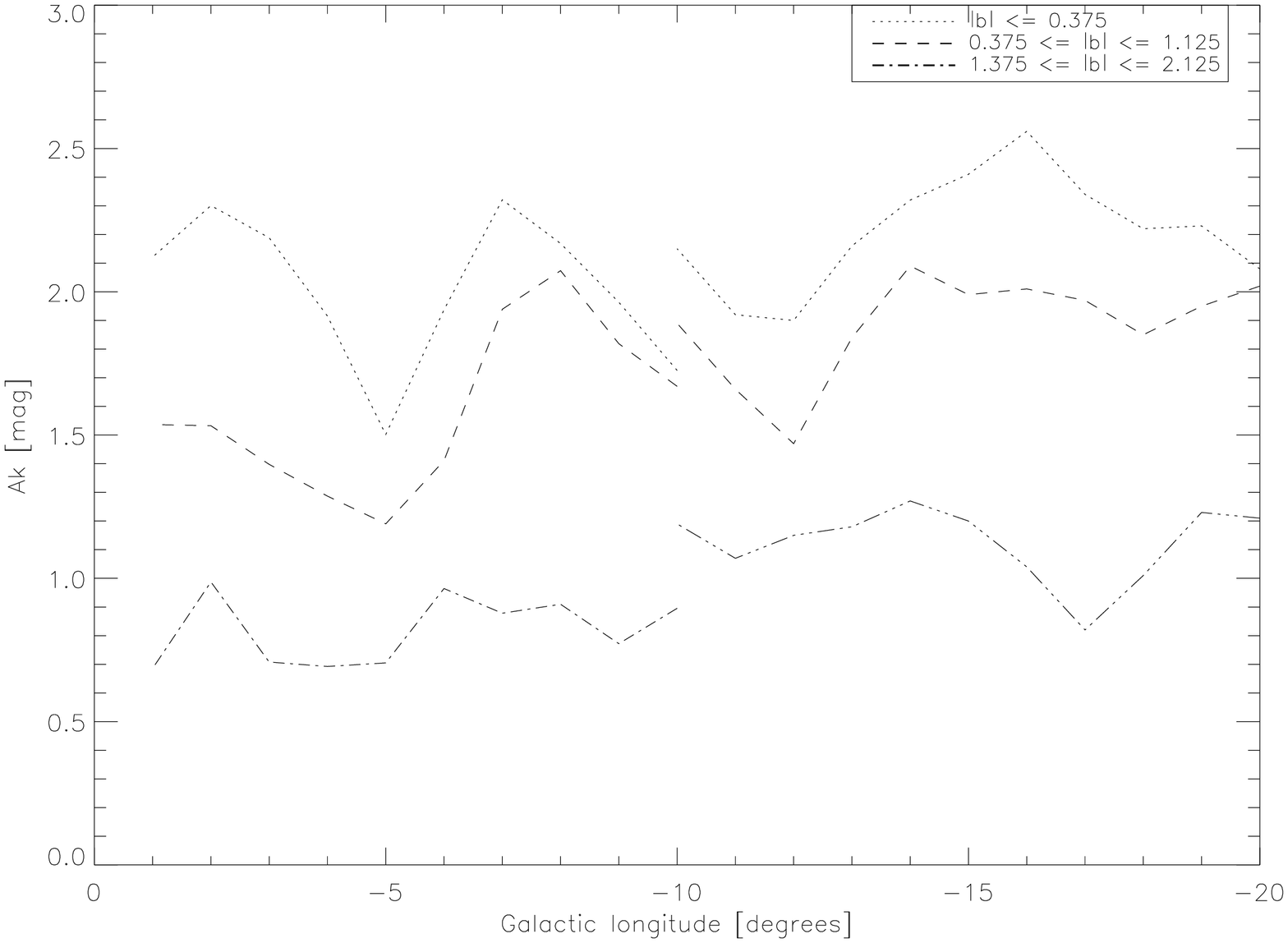}}
\resizebox{9.0cm}{!}{\includegraphics[angle=90,scale=.90]{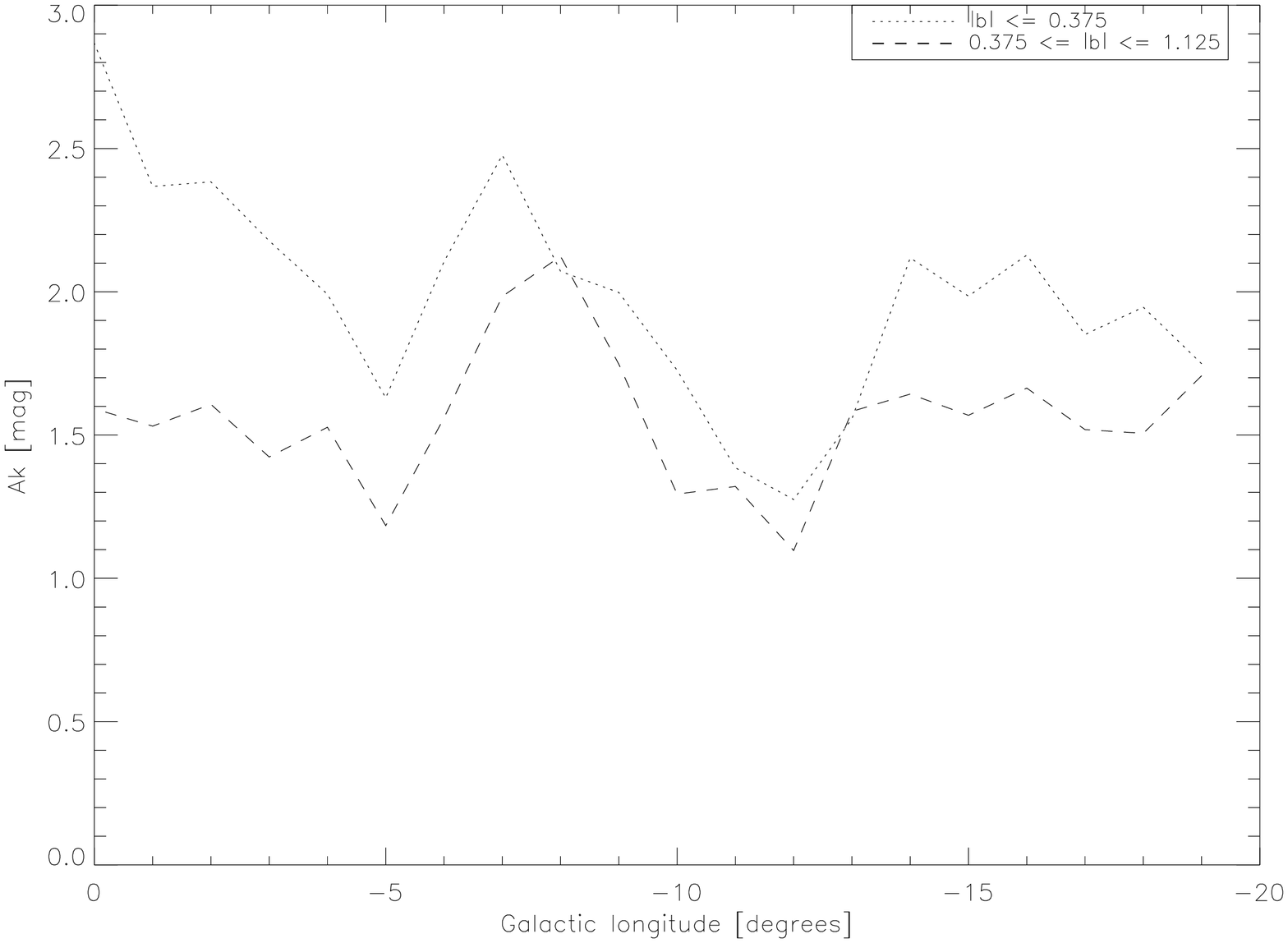}}
\caption{Longitudinal profile for three ranges of latitude as
mentioned in the legend. Upper left panel: star counts (for the
observed counts up to $K_{\rm  s}=$ 13.5 mag) for magnitude
corrected by extinction effects; upper right panel: relative
difference between counts with and without extinction correction.
lower left panel: \textit{Averaged extinction} predicted by
Gonz\'alez et al. (2012) and Marshall et al. (2006) interstellar
extinction model, for Galactic longitudes: $-10^\circ < \ell <
0^\circ$ and $-20^\circ < \ell < -10^\circ$, respectively; lower
right panel: \textit{Averaged extinction} predicted by Nidever et
al. (2012). In figures, we have considered a median within one
degree in longitude.}
\end{figure*}

Another visible feature is the end of the bulge loss revealed by
the change of slope in the decrease in star counts at $\ell \sim
-8^\circ$ for the higher latitudes. For instance for $-8.0^\circ <
\ell \leq -3.0^\circ$, the counts have an average value equal to
$\sim$ 1.5 x $10^{5}$ to 1.1 x $10^{5}$ for $\ell \sim -8^\circ$.
Also, it can be seen that counts for the mid latitude
(0.375$^\circ < |b| \leq 1.125^{o}$) are greater than for the low
latitude ($|b| \leq 0.375^\circ$) for some longitude ranges
($-5.0^\circ < \ell < -2.5^\circ$ and $\ell > -16.5^\circ$). This
latter feature can be accounted for by errors in the extinction
correction and the shift in the stellar distribution centroid (see
Section 4.2).

A \textit{rough visualization} of the interstellar extinction
toward the Galactic bar and bulge can be obtained by displaying
the ratio between the counts with and without extinction
correction. Figure 3 (upper right panel) depicts the variation in
relative counts with longitude, which also provides a rough
estimate of the variation in interstellar extinction. To produce
this estimate, we have considered the relative difference between
the corrected and uncorrected counts for each interval of one
degree in longitude. The interval of Galactic latitude is shown in
the legend of figures.

To compare the interstellar extinction values obtained in the
present work with the obtained ones by other works, we have
considered three different estimates covering two ranges of
Galactic longitude: i) the map produced by Gonz\'alez et al.
(2012) using new and deep VVV data and red clump method covering
($-10^\circ < \ell < 0^\circ$); ii) the 3D interstellar extinction
model proposed by Marshall et al. (2006) using both the
Besan\c{c}on Galaxy Model (Robin et al. 2003) and 2MASS data. In
this last model, we used as extinction the value that corresponds
to the maximum value of distance, ranging from 10 to 13 kpc, for
Galactic longitudes covering the regions located at $-20^\circ <
\ell < -10^\circ$; iii) the maps provided by Nidever, Zasowski \&
Majewski (2012, hereafter Nidever et al. 2012) based on the
Rayleigh-Jeans color excess method using GLIMPSE$-I$,$-II$,$-3D$
data.

The maps of Gonz\'alez et al. (2012) and Nidever et al. (2012)
were available during the referee process, as well as the 3D
extinction model provided by Chen et al. (2013). We have not used
this last one in our comparison since, as pointed out by Chen et
al. (2013), the results for the region of present study are in
good agreement with ones obtained by Gonz\'alez et al. (2012).

It should be noted that each point of extinction obtained by the
method provided by other authors was obtained considering the
median of extinction for an interval of one degree in Galactic
longitude. For comparison, Figure 3 (lower left panel) displays
the average interstellar extinction distribution provided by
Gonz\'alez et al. (2012) and Marshall et al. (2006) for the same
latitude ranges and adopting the median extinction for each
pointing separated by one degree in longitude. As can be seen in
Figure 3 (lower left panel), \textit{averaged} extinction in the
Galactic plane reaches about A$_V = 25$ mag with most values
ranging from 15 to 20 mag. Generally, our \textit{averaged}
extinction variations in relative star counts are in agreement
with those provided by Gonz\'alez et al. (2012) and Marshall et
al. (2006).

Since the star counts analysis is based on magnitude limited
samples, an overestimated absorption will lead to overestimate
star counts, because if the estimated extinction is too large,
then stars that would fall beyond the magnitude cut-off would be
overcorrected and thus be counted, biasing the \textit{corrected}
counts upward. Another effect is crowding. Since in-plane fields
are more affected by crowding, here we will primarily lose the
dimmer and redder stars than the ones that yield the high
extinction values in off-plane fields.

This context provides an explanation for the awkwardly lower
counts of the lowest latitude sample ($|b| \leq 0.375^\circ$) in
some ranges ($-5.0^\circ < \ell < -2.5^\circ$ and $\ell >
-16.5^\circ$), where the disk is expected to have higher counts,
compared to the intermediate latitude sample ($0.375^\circ < |b|
\leq 1.125^\circ$). Comparison with the interstellar extinction
distribution proposed by Gonz\'alez et al. (2012)  gives a hint
that at $-5.0^\circ < \ell < -2.0^\circ$ the lower latitude
extinction is underestimated, and the intermediate latitude range
extinction is overestimated, giving rise to an extinction peak.
Both effects naturally account for the apparently lower star
counts of the lower latitude region compared to the mid-latitude
region. The same reasoning applies to $\ell < -16.5^\circ$.

In any case, a significant increase in extinction for Galactic
longitudes beyond (in the negative sense) $\ell \sim -14^\circ$ is
appreciated for the three latitude ranges and are stronger for
$|b| \leq 0.375^\circ$. The extinction grows until $\ell \sim
-16^\circ$, decreasing towards $\ell \sim -20^\circ$ for the low
latitude region, but shows a slight increase for the higher
latitudes, see the discussion and references therein of Calbet et
al. (1996) and Gonz\'{a}lez-Fern\'{a}ndez et al. (2012).

The same feature was observed by LC01 in the regions towards long
Galactic bar while comparing DENIS and CAIN data (their Figure 8).
The same authors also noticed the extra extinction from $\ell \sim
-12^\circ$ to $\ell \sim -8^\circ$. The deeper VVV photometry in
the present work allows a better definition of the feature,
clearly displayed by very high extinction observed from $\ell \sim
-11^\circ$ to $\ell \sim -6^\circ$.

In general the maps (Figure 3, lower right panel) provided by
Nidever et al. (2012) presents a good agreement with obtained ones
by Gonz\'alez et al. (2012), as well as with the one obtained by
us. Some differences can be seen, such as the valley observed for
Galactic latitudes $|b| \leq 1^\circ$, in our case for $\ell \sim
-13.0^\circ$ and for $\ell \sim -12.0^\circ$ (Nidever et al. 2012)
This can be attributed to the averaged extinction process used to
elaborate the figure or even to the fact on using deep VVV data.
To obtain the extinction from Nidever et al. (2012), we used the
option \textit{all} for \emph{stellar population} and
\textit{percentil} for \emph{map type}. The maps of Nidever et al.
(2012) are available at their web page\footnote
{http://www.astro.virginia.edu/rjce/}.

Another point that should be considered in the comparison of our
extinction method with the obtained ones by other authors is the
point that we do not use local sources with $J-K_{\rm s} < 0.5$,
or the sources with $0.5 < J-K_{\rm s} < 1.0$ have no extinction
correction applied to them. Despite all these points, we can see
good agreement with our \textit{extinction} estimate with other
author. Detailed discussion about interstellar extinction
distribution along the Galactic bar as well as analysis of 3D
distribution is beyond the scope of the present paper.

It is important to mention that an aspect should be considered
when comparing our data with the used ones by other author, for
instance with Nidever et al. (2012, Figure 1c), which is
applicable to ($\ell$,b) = (42$^\circ$, 0$^\circ$). The point is
that for this direction, we are mainly observing disk stars, with
a maximum density at distance of 6 kpc from the Sun. This makes
the population at low distances of Sun predominant at $K_{\rm
s}=12$, with a higher ratio of dwarfs, which are bluer than
$(J-K_{\rm s})_{0}= $ 1.0. This is not our case (see
color-magnitude diagrams in Paper I).

\subsection{Latitudinal counts}

Latitudinal star counts profiles provide valuable information on
the structure of the Galaxy (Freudenreich et al. 1994). These
profiles allow to  infer the thickness of the dust-gas and stellar
layers and their displacement with respect to the Galactic plane,
among other applications. At the longitudes toward the Galactic
anti-center, these profiles allow characterization of structures,
such as the Galactic warp, among others. At the longitudes close
to the Galactic center, they enable determining bar and bulge
morphological parameters. Latitudinal profiles towards the long
Galactic bar were built for the VVV+2MASS data, using extinction
correction determined in this work (Section 3.1).

Gaussians were fitted (see some examples in Figure 4) to the
observed latitudinal profiles limited to $K_{\rm  s} \leq 13.5$
mag and to colors $J-K_{\rm s} \geq$ 0.5 mag. The fits were done
using an IDL version of a genetic algorithm called PIKAIA
(Charbonneau, 1995). Given that there are three parameters to be
determined from a small volume of data, a small number of
generations (100) with 60 populations each were adopted to insure
a faster convergence. Twenty independent runs were performed, and
the final solution is the median for the runs with $\chi^{2}$
within 1$-\sigma$ of the distribution. The errors bars were
determined from the standard deviation of the parameters. The
merit function (a similar one was used by Larsen \& Humphreys,
2003) used was

\bigskip
\begin{equation} \label{equ:xi}
\displaystyle \chi^{2} =
\sum_{i=1}^n(N_{i}{(obs)}-N_{i}{(model)})^{2}/N_{i}{(obs)},
\end{equation}
\noindent
\bigskip

\noindent where $N_{model}$ and $N_{obs}$ are the modeled and
observed counts for each bin of Galactic latitude ($i$).

It was obtained by adjusting the three parameters of the Gaussian
function:

\bigskip
\begin{equation} \label{equ:rjce}
\displaystyle f(z)= A \times exp(-|z-z_c|^{2}/K),
\end{equation}
\noindent
\bigskip

\noindent in which $K=2 \sigma ^2$ is related to thickness,
\emph{A} is the central surface density of the stars, and $z_c$
the offset from the midplane. The range of parameters are
presented in Table 2.

\begin{figure}
\includegraphics[width=8.1cm]{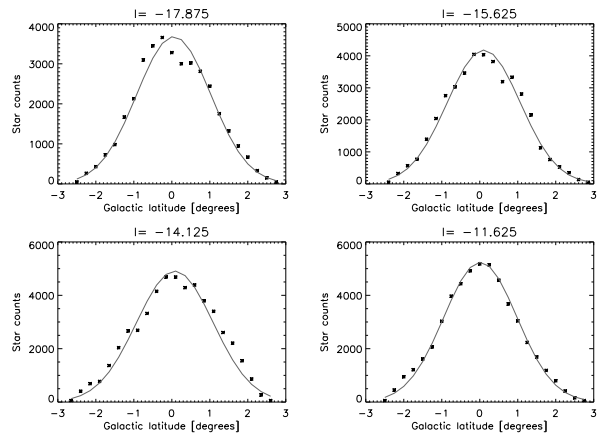}
\centering \caption{Examples of Gaussian adjustments (lines) for
the latitudinal profiles of the counts in $K_{\rm s}$ (asterisks)
corrected by interstellar extinction effects. Galactic longitude
is indicated at the top of the panels.} \label{figure4}
\end{figure}

As proposed by Cabrera-Lavers et al. (2007a), we also tried to
adjust a $sech^{2}$ function; however, Equation (7) produced a
better fit to our data. For the VVV+2MASS set, splitting it in
0.25$^\circ \times 0.25^\circ$ regions gives 81 pointings with
different longitudes, each one with 22 different latitude
pointings. This sampling allows robust fits with good spatial
resolution.

Figure 5 shows the variation in Gaussian centroids derived from
the latitudinal profile fitting VVV+2MASS data. The variation is
almost flat from $\ell \sim -11.5^\circ$ to $\ell \sim -10^\circ$,
reaching a local minimum at $\ell \sim -11.5^\circ$ and increasing
linearly to $\ell \sim -12.75^\circ$. After that, a decrease is
seen until reaches a minimum at $\ell \sim -13.8^\circ$ (z$_c =
-0.05^\circ$). A significant increase is then seen reaching a
maximum at $\ell \sim -14.5^\circ$, after that there is a small
local decrease, and the centroid then remains roughly constant
until $\ell \sim -18.5^\circ$.

\begin{figure}
\includegraphics[width=7.0cm,angle=90]{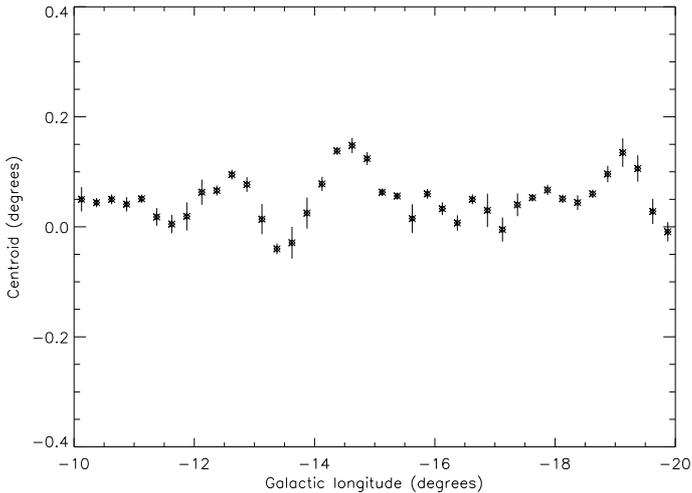}
\centering \caption{Longitudinal profile for the stellar centroid
variation obtained from adjustments of the latitudinal profiles of
the counts $K_{\rm  s}$.} \label{figure5}
\end{figure}

We have restricted our analysis up to longitudes lower than $\ell
\sim -10^\circ$ since many latitudinal profiles tend to have
irregular shapes toward the bulge, which removes meaning from
Gaussian fits. These irregularities could be due to inadequate
interstellar extinction correction, or more likely, to VVV
photometric incompleteness toward the more crowded Galactic center
region. This is supported by the study of Am\^{o}res et al. (2012)
aimed at detecting galaxies in the VVV database. The authors
report that a closer examination reveals that many sources
classified by VVV as extended objects are in fact composed of
systems with two or three stars.

\bigskip
\begin{table}
\centering \caption{Range of parameters used to adjust the
function represented by Equation (7). x$_1$, x$_2$, and x$_3$ are
the normalization factor used by PIKAIA, ranging from 0.0 to 1.0.}
\begin{tabular}{cc}
\hline parameters & range \\
\hline \noalign{\smallskip}
$z_c$ &  $-3.0 + 9.0x_1$\\
width &  $12.0x_2$   \\
heigth&  maximum data height $\times$ 1.3 $\times$ x$_3$    \\
\noalign{\smallskip} \hline
\end{tabular}
\end{table}
\bigskip

Freudenreich et al. (1994) analyzed latitudinal profiles for
COBE/DIRBE emission, finding warp and flare structures toward the
Galactic anti-center at 240 $\mu$m, which is dominated by dust
emission. At near and mid infrared DIRBE/COBE wavelengths one can
see the displacement in latitude of the brightness peak toward the
Galactic center, but the resolution of their latitudinal profiles
(each 10$^\circ$) does not allow the bar region to be identified.
On the other hand, the adjustment for the latitudinal profiles
obtained by BL91 (their Figure 12) shows clearly a tilt in the
variation of the centroid of the surface brightness distribution,
with a typical deviation of $\pm 0.4^\circ$ in latitude;
unfortunately, their figure only shows the variation from Galactic
longitudes $(|\ell| < 10^\circ)$.

By adjusting data from the IRAS 100 $\mu$m band that traces dust
distribution (AL05) at one degree interval (Am\^{o}res 2005),
found a centroid distribution compatible to obtaining one by BL91
with a significative decrease for longitudes lower than $\ell \sim
-13^\circ$, as also found in the present work with VVV data. In
fact, the asymmetric shape for the stellar distribution that can
be seen in Figure 5 suggests a similar feature for the dust
distribution that dominates this region. Am\^{o}res (2005) found a
variation of 0.3 degrees in $z_c$ that is compatible with the
expected scenario for regions of leading dust lanes. The author
also analyzed latitudinal profiles for HI for specific ranges of
velocities (belonging to peaks related to FIR emission) and a
significant variation in the centroid is found toward $\ell \sim
-15^\circ$. Making the assumption of non-circular movements at the
Galactic center, a distance equal to 10.7 kpc (considering v
$\sim$ -110 $kms^{-1}$) was found. Marshall et al. (2008) analyzed
the latitudinal profiles for extinction obtained with the
extinction model proposed by Marshall et al. (2006) and CO but for
a longitude range of $10^\circ$ to $-4^\circ$.

From the latitudinal profile fits we obtained a parameter that we
call to \emph{K} (see Equation 7), which is also useful in
estimating the vertical thickness of the bar that can be estimated
using the expression

\begin{equation} \label{equ:a8mu}
\displaystyle h_z=d\sin (\sigma),
\end{equation}
\noindent

\noindent where $d$ is the \emph{average distance} for the points,
e.g., $(\ell, b)$ pairs considered in this work to the far side of
the Galactic bar, equal to approximately 10.0 kpc (LC01). We get
an average $\sigma =1.01\pm 0.03$ degrees; hence we get a vertical
thickness approximately equal to 176$\pm $6 pc. This value is
higher than 100 pc obtained by Cabrera-Lavers et al. (2007) using
red clump stars. First, we must take into account that we are
using a wider range of stellar types than red clumps, and we are
also observing the bar with a different angle (at positive
longitudes, we observe the bar almost perpendicularly, whereas at
negative longitudes our line of sight is more tangential to the
bar). More importantly, in our star counts variations, we also
include variations from the disk+bulge+bar instead of the isolated
bar, therefore we get a higher thickness which is the average of
the three embedded structures with their corresponding weights. A
rough estimation of the bar thickness might be given considering
that $\sigma =1.01\pm 0.03^\circ$ is the $\sigma $ of the sum of
two Gaussians.\footnote{If we have a Gaussian for the bar with
$\sigma_{bar}=$0.7 and a second Gaussian for the disk+bulge with
$\sigma _{b+d}$=2.0 with the same amplitude, the sum of both
functions approximately gives a Gaussian with
$\sigma_{total}=1.01$.}. We know that the contribution of the
disk+bulge gives a $\sigma _{b+d}\approx 2.0^\circ$ at $\ell \sim
-12^\circ$ (L\'opez-Corredoira et al. 2004, 2005), and roughly
bulge+disk gives the same counts as the bar (see Fig. 3, and
compare the inplane counts with counts at $b=2^\circ$); hence,
$\sigma _{bar}=$ 0.7$^\circ$, so the vertical thickness of the bar
alone should be around 120 pc, closer to Cabrera-Lavers et al.
(2007) results. We state that bulge+disk have roughly the same
counts at $\ell=12^\circ$, b=0$^\circ$ as the bar, because at
b=1.75$^\circ$ (where the bar contribution is negligible), they
are 2.5 times lower (instead of 1.46 times lower expected from the
Gaussian of bulge+disk) attributing the difference to the bar.

\section{Discussion}

\begin{figure*}[t]
\centering
\resizebox{15.0cm}{!}{\includegraphics[angle=0,scale=1.0]{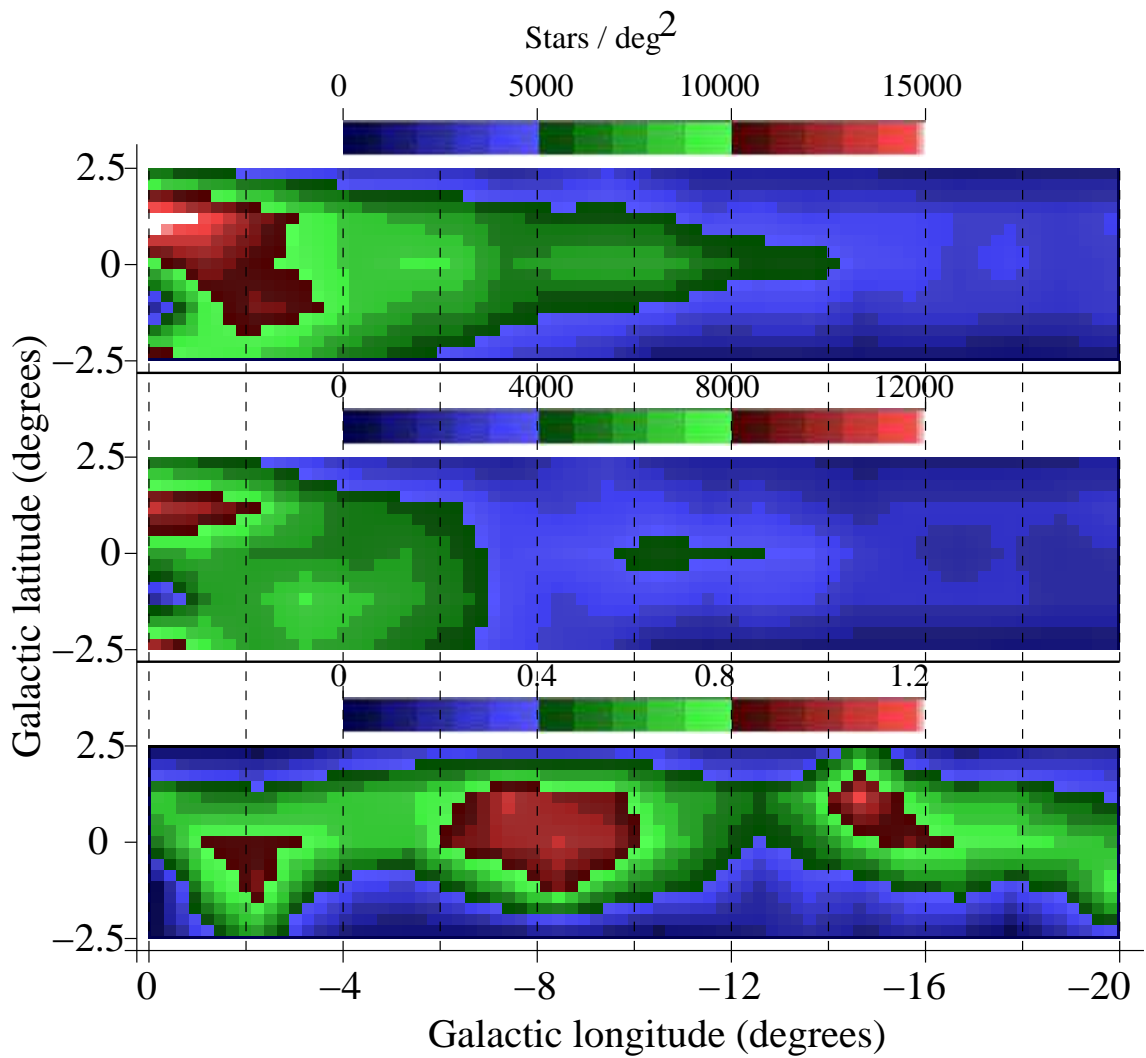}}
\caption{VVV counts (up to $K_{\rm  s}=$ 13.5 mag) towards the
long Galactic bar: corrected for interstellar effects (upper
panel); without correction (middle panel); relative difference
(lower panel), star counts binned ($\Delta l=\Delta b=1^\circ $)
and interpolated.}
\end{figure*}

Figure 6 shows the counts up to $K_{\rm  s}=$ 13.5 mag for the
long bar region with and without extinction correction, as well as
for the residual differences. In Figure 6 (upper panel), a
structure ending toward $\ell \sim -8^\circ$ is seen that can be
attributed to the Galactic bulge. A long bar structure can be seen
as the green region extending up to $\ell \sim -14^\circ$. For
longitudes after $\ell \sim -15^\circ$, the stellar density
decreases as illustrated by the lighter blue area in Figure 6
(upper panel). In this figure, part of stars are red giants. The
far side of the Galactic bar as presented in this paper has an end
at $\ell \approx -14^\circ$, and this is not only due to the high
extinction observed after bar ends but also the density of stars
that falls considerably in this longitude as seen from several
surveys in the NIR and MIR.

Moreover, works on Galactic resonances (Amaral \& L\'{e}pine,
1997; L\'{e}pine et al. 2001; Mishurov, Am\^{o}res \&  L\'{e}pine
2009 and references therein) explain how regions toward them, such
as the end of the Galactic bar, suffer the effects of corotation.
In these regions, there is also smaller quantity of gas and stars.
Briefly, the effects of these resonances can be understood as
follows. The gas dynamics in the perturbed potential of the spiral
arms is such that inside the corotation radius, a net flow is
produced toward the center. Beyond corotation, a net flow is
produced toward the external parts of the disk. This dynamics
results in a pumping gas out from the corotation region
(Am\^{o}res, L\'{e}pine \& Mishurov, 2009).

Figure 6 (middle panel) clearly illustrates the importance of
correcting for extinction. Without it, the structure identified in
Figure 6 (upper panel) is not observed. The star counts decrease
too quickly at $\ell \sim -8^\circ$, and the structure is too
thick for $\ell > -8^\circ$ and too thin for $\ell < -8^\circ$.
Furthermore, the counts without extinction correction produce a
shape without the symmetry seen in Figure 6 (upper panel). In
Figure 3, we pointed that there is a clear decrease in the counts
until $\ell \sim -14^\circ$. Certainly, the smoothing in longitude
and latitude decreases the contrast in Figure 3, and this is
better observed in Figure 6 (upper panel).

Figure 6 (lower panel) shows a map for the relative difference
(see Section 4.1) in counts between the upper panels, which
roughly corresponds to an extinction map. In this figure, the high
extinction structures toward the end of the long Galactic bar are
highlighted. LC01 argue that longitudes corresponding to the
negative limit of the bar are at almost twice the distance
relative to the center as those on the nearest side, which are
also affected by higher extinction. This feature can also be seen
in the comparison with GLIMPSE data shown in Figure 7.

\begin{figure*}[t]
\centering
\resizebox{14.0cm}{!}{\includegraphics[angle=0,scale=.90]{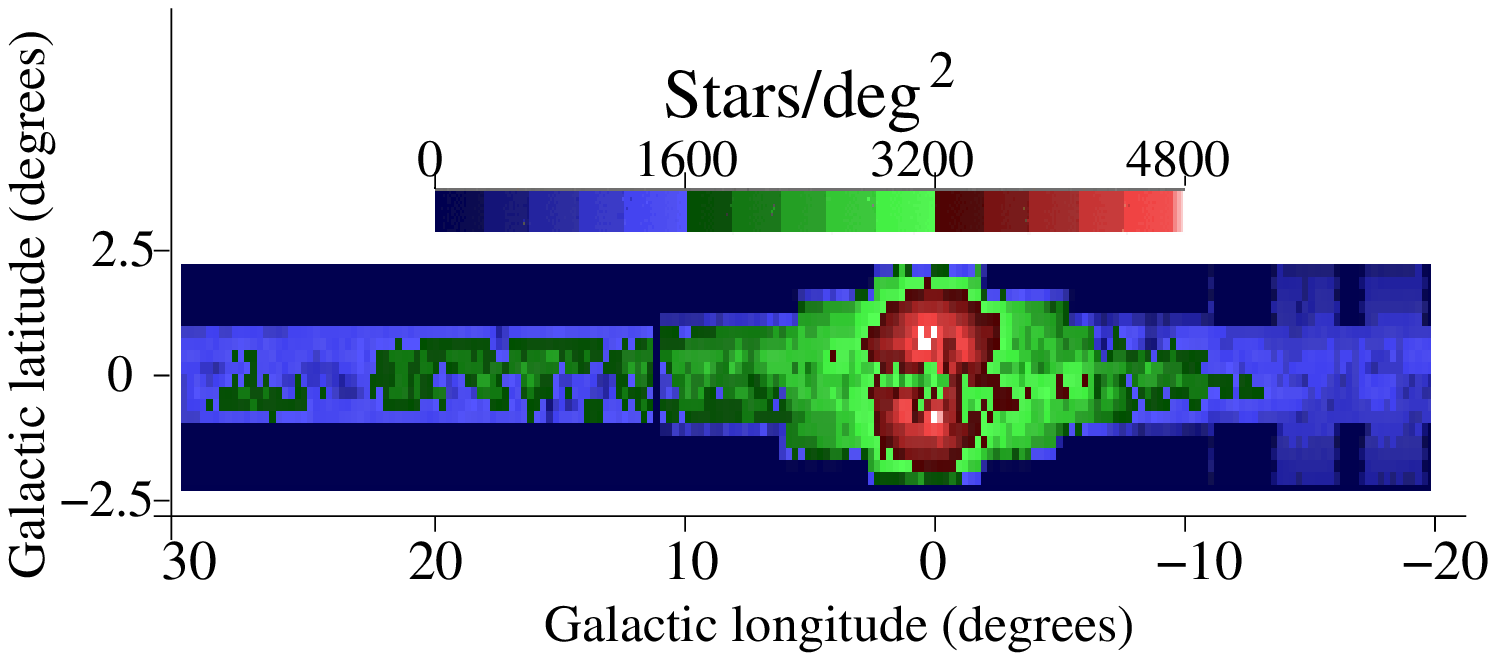}}
\resizebox{14.0cm}{!}{\includegraphics[angle=0,scale=.90]{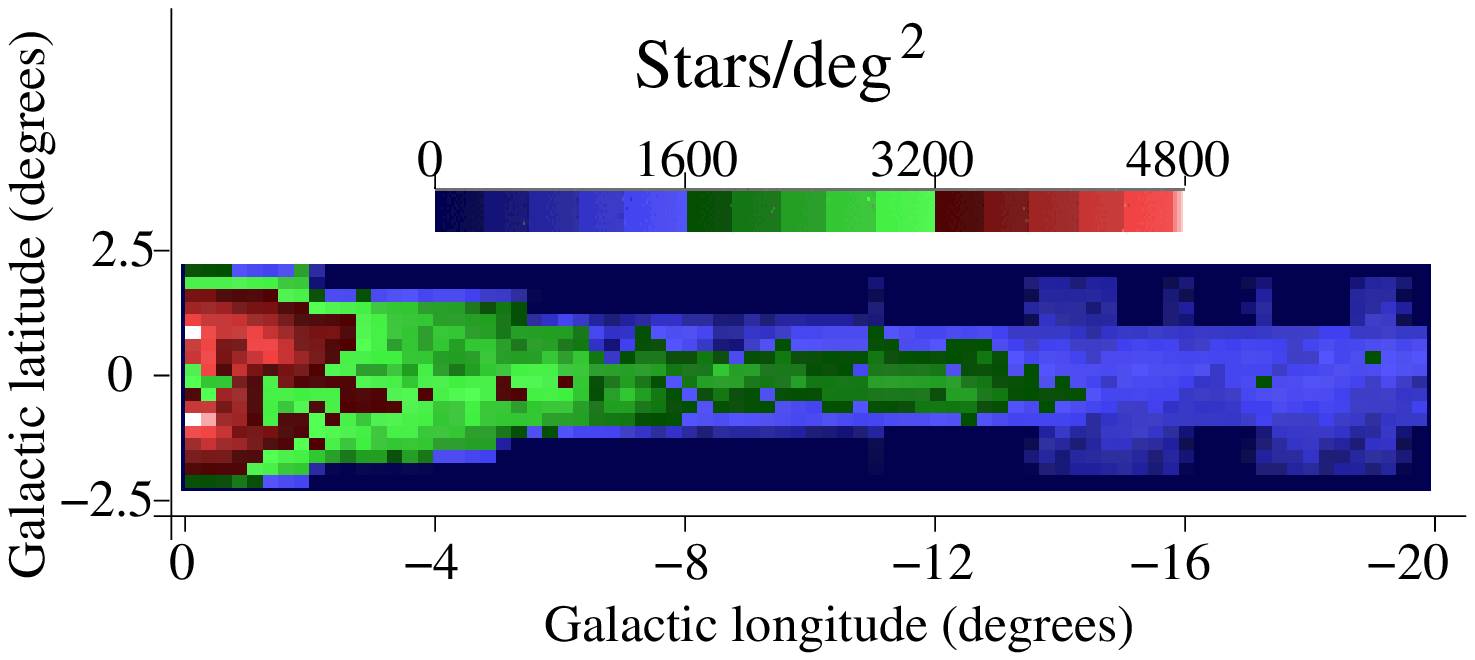}}
\resizebox{14.0cm}{!}{\includegraphics[angle=0,scale=.90]{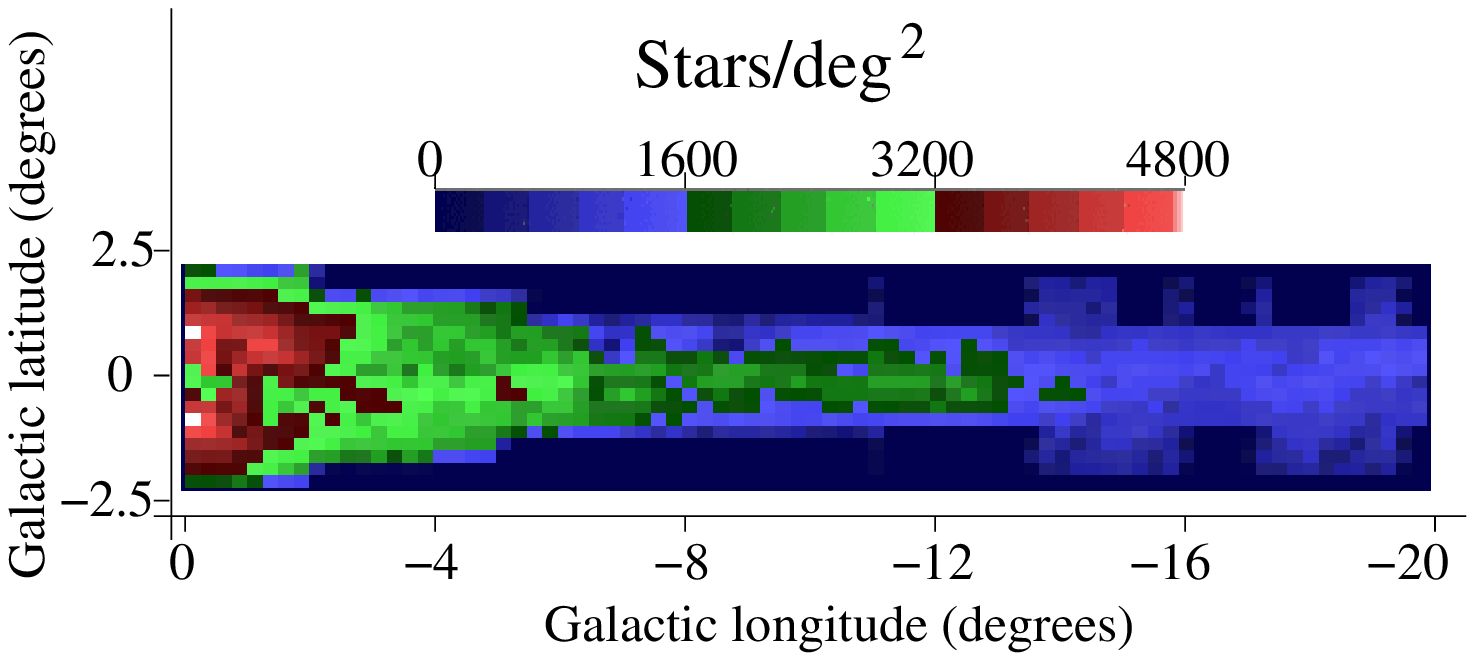}}
\caption{GLIMPSE counts (up to [8.0] $\mu$m = 11.0 mag) using
extinction method presented in Equation (4) that uses the combined
data (upper and middle panel). Upper panel: GLIMPSE+2MASS. Middle
panel: GLIMPSE+VVV+2MASS with interstellar extinction correction
going deeper in the comparisons as presented in Table 2. Lower
panel: GLIMPSE+VVV+2MASS without interstellar extinction
correction.}
\end{figure*}

The difference between $m_{K_{\rm s}}$ and $m_{K_{\rm s},{\rm
correc}}$ is proportional to the average extinction (see Equation
1); there is only dependence on the structure distance through the
distribution of dust and stars along the line of sight. Therefore,
Figure 6 (lower) is a map of average extinction until the distance
of the average position of the stars along the line of sight. It
is also possible to identify an asymmetry in the distribution of
interstellar extinction with respect to the formal Galactic plane
from $\ell \sim -16^\circ$ to $\ell \sim -14^\circ$. This feature
was also identified in IRAS 100 $\mu$m (Am\^{o}res 2005).

As found by LC01 (and discussed in Section 4.1), we also identify
high extinction toward $-10^\circ < \ell < -8^\circ$ in our
VVV+2MASS data. The extinction decreases until the bar ends at
$\ell \sim -14^\circ$,  which can be interpreted as the dust lanes
mentioned by Calbet et al. (1996), Rodriguez$-$Fernandez et al.
(2006), Nagai et al. (2007), Marshall et al. (2008), and Paper I.
Then, a high extinction region extends up to $\ell < -19^\circ$ in
the proximity of the 3 kpc-arm. This latter high extinction
feature contrasts with the star counts density in Figure 3 (upper
right panel).

Concerning the near side of the Galactic bar, several works have
placed it at $\ell \sim +27^\circ$, approximately 5.7 kpc from the
Sun, as pointed out by Hammerseley et al. (1994, 2000, e.g. H94,
H00), Garz\'{o}n (1993),  LC01, Cabrera-Lavers et al. (2007a), and
Picaud et al. (2003), among others.

We have adopted distance to the far side of Galactic bar of 11.1
kpc (as also used by LC01). We also argue that the long bar tip on
the far side is at longitude limit of $\ell \approx -14^\circ $.
If we consider a semi-major axis length L$_{0}$ of 4.0 $\pm$ 0.5
kpc, values that are compatible with the ones obtained by LC01,
H00, H94, among others, we can estimate the inclination angle
($\alpha$) using

\bigskip
\begin{equation} \label{equ:a8mu}
\displaystyle \alpha = \ell + asin{[(R_{0}/L_{0})sin(\ell)]},
\end{equation}
\noindent
\bigskip

\noindent where $R_{0} =$ 8.0 kpc is the distance from the Sun to
the Galactic center, $L_{0}$ is the semi-major axis length and
$\ell$ is the longitude that corresponds to the end of bar.

We can obtain $\alpha$ by considering the error for both $\ell$
and $L_{0}$. We assume that $\ell \approx -14^\circ $ and
$L_0=4.0\pm 0.5$ kpc. Using Equation (9) we obtained $\alpha
=43^\circ\pm 5^\circ$. This angle is the same one as the measured
values in the positive longitudes (e.g. H00), so we conclude the
bar must be straight.

This value agrees with other authors such as Peters (1976), Nakai
(1992), Hammersely et al. (2000), and Sevenster et al. (1999), who
report 45$^\circ$, 43$^\circ$, and 45$^\circ$, respectively. This
value also agrees with findings pointed out by LC01 (43.0$^\circ$)
and later by several authors, e.g., LC07 and references therein;
Cabrera-Lavers et al. 2007b; Benjamin et al. (2005); Vallenari et
al. (2008) report an angle of 45$^\circ$. A schematic view of bar
representation can be seen in Figure 8 of Paper I. Figure 7
(middle panel) shows the counts at [8.0] $\mu$m with a cut-off at
11.0 mag and corrected for extinction using the relations provided
by M11. Determination of the extinction correction made use  of
the deep combined VVV+2MASS+GLIMPSE data set, as described in
Section 2.2.

A structure, coded in darker green and seen reaching $\ell \sim
-13^\circ$, corresponds to the long Galactic bar and is then
followed by a region having smaller and almost constant star
counts. As in Figure 6, the bulge is clearly seen, ending toward
$\ell \sim -8^\circ$. Figure 7 (upper panel) shows the map using
the extinction correction provided by M11 using only the less deep
2MASS+GLIMPSE combined data (no VVV).

Benjamin et al. (2005) used GLIMPSE data at [4.5] $\mu$m, to
derive an inclination angle of 44 $\pm 10^\circ$ for the bar, with
$R_{bar} = 4.5 \pm$ 0.5 kpc. As pointed by LC07 their general
results agrees with those obtained by LC01, also claiming for a
long bar. Their longitudinal source-density profiles at [4.5]
$\mu$m show a decrease in the counts at $\ell \sim -13^\circ$.
They also produced a star counts map in which one can identify
some structures similar to those reported in this work. However,
their picture is not as sharp as our Figure 7 (middle panel),
which we think is due to our improved extinction correction using
deep J and $K_{\rm  s}$ photometry and to extinction being lower
at [8.0] $\mu$m. A marked difference between both maps is that
Benjamin et al. (2005) found significantly fewer stars at Galactic
longitudes from $\ell \sim -17^\circ$ to -15$^\circ$, a region
that we point out as a high extinction zone.

In Figure 2 of Nidever et al. (2012), there is some gap after
$\ell \sim -15^\circ$, although it is not very clear, probably
because 2MASS data is not deep enough on the range of magnitudes
used. However, the features of the Galactic bar can be seen in
GLIMPSE data, when observed considering appropriate range of
magnitudes from $\sim 14$ to $\sim 6.5$, see for instance maps in
Churchwell et al. (2009) and Benjamin et al. (2005) as mentioned
in the last paragraph.

Figure 7 (upper panel) shows the counts at [8.0] $\mu$m but only
based on 2MASS photometry instead of combined VVV+2MASS data. It
can be seen that the structure of the long bar ends towards $\ell
\sim -11^\circ$, unlike the limit obtained when based on the
combined data, thus allowing for a deeper analysis of the Galactic
bar. Figure 7 (lower panel) shows the counts [8.0] $\mu$m that are
almost similar to the one presented in Figure 7 (middle). The
similarity between the counts highlight that extinction is much
lower at [8.0] $\mu$m, as discussed in Section 3.2 (see Equation
4).

\begin{figure}
\includegraphics[scale=0.34,angle=90]{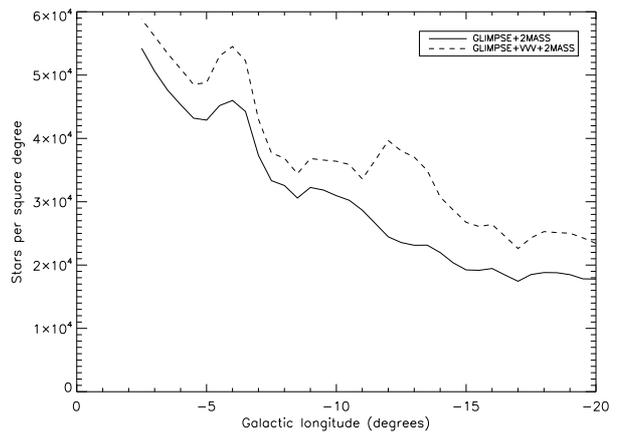}
\centering \caption{Longitudinal profile for GLIMPSE counts in
[8.0] $\mu$m using the extinction method presented in Equation (4)
that uses the combined data.}\label{figure8}
\end{figure}

This can be visualized better in Figure 8, which shows
longitudinal profiles for counts at [8.0] $\mu$m using only
GLIMPSE+2MASS and combined GLIMPSE+2MASS+VVV data. There are two
main aspects to be considered by this figure: i) the number of
sources clearly increases with the addition of VVV data since
there are more matches to GLIMPSE sources that were not detected
by 2MASS (see Table 1); ii) some differences in the shape of
profile (most for $\ell <$ -10$^\circ$) can also be noted and
attributed to the use of extinction correction with more stars.
The relative count differences attributed by combining the VVV
data are approximately 20\% and 27\% for $\ell \ge$ -10$^\circ$
and $\ell <$ -10$^\circ$, respectively. A structure with a double
peak can be seen within the longitude range: -14$^\circ < \ell
<-12^\circ$. For $\ell <$ -14$^\circ$ a significant decrease in
the counts was independently found by LC07 using MSX data at [8.7]
$\mu$m.

Concerning another study of star counts toward the far side of
long bar; Benjamin et al. (2005) found using GLIMPSE at [4.5]
$\mu$m, approximately 2.0$\times10^{4}$ sources per square degree.
LC01 find counts for $m_{K} \leq$ 9.0 of approximately 2,000 using
DENIS data averaged over $\Delta\ell=$ 5$^\circ$. We can estimate
the counts expected by a model, in this case using a similar model
as proposed by LC01 with disk density given by
L\'{o}pez-Corredoira et al. (2002, 2004), with the luminosity
function proposed by Eaton et al. (1984) and the counts in the
Galactic bar based on the density distribution proposed by
L\'{o}pez-Corredoira et al. (2007) with the luminosity function
considered as an average of the disk and bulge luminosity
function. We have obtained counts in good agreement with ones
observed by VVV. For instance, for $\ell$ $\sim -15.0^\circ$ for
two ranges of Galactic latitudes, 0.375$^\circ < |b| \leq
1.125^\circ$ and 1.375$^{o}$ $\leq |b| \leq 2.125^\circ$, we
estimate $10^{4}-10^{5}$ sources per square degree (considering as
magnitude limit $K_{\rm  s}$ = 13.5 mag), respectively. The values
of parameters used in the model are in good agreement with those
ones obtained in this work.

\section{Conclusions}

In this paper we have used deep VVV photometric data to perform a
detailed analysis of the far side of long Galactic bar, which is
the less studied side, by far, in the literature. This analysis
was based on star counts in the $K_{\rm s}$ band for the region
extending from $\ell \sim -20.0^\circ$ to 0.0$^\circ$ and for
Galactic latitudes $|b| \leq 2^\circ$.

We used the extinction correction recipes provided by
L\'opez-Corredoira et al. (2001) and Majewski et al. (2011) on
combined VVV+2MASS and GLIMPSE data, respectively. Since there is
more extinction toward negative longitudes than toward positive
longitudes (LC01), the average extinction correction methods
employed in this work has the advantage of using many distant
stars with high J-$K_{\rm s}$ and H values, as provided by VVV
survey.

Even though the method used in this work had been presented and
used in other works, we expanded it by comparing it with other
more complexes extinction models, as well as with star counts
simulations as presented in Appendix A.

The use and combination of extinction methods with VVV data allow
us to better account for the effects of interstellar extinction on
star counts for both the $K_{\rm s}$ and [8.0] $\mu$m bands. An
extinction map was produced (Figure 6, lower panel), revealing
high extinction regions along the bar, more notably slightly
farther than the bar end ($\ell \sim -14^\circ$). A tilt in the
dust distribution (Figure 6, lower panel) is noted, observed
towards $\ell \sim -15^\circ$ ($b \sim 0.5 ^\circ$) with similar
characteristics to those observed at IRAS 100 $\mu$m (Am\^{o}res,
2005), which is a feature not observed, for instance, in the
Marshall et al. (2006) extinction maps.

Even though this work was not devoted to the study of interstellar
extinction, our results highlight the potential use of the VVV
survey for modeling the dust distribution for both 2D and 3D
models as presented by Gonz\'alez et al. (2012) and Chen et al.
(2013). This type of analysis also has the potential for studying
interstellar extinction in the Galactic disk with the use of NIR
and MIR colors using combined VVV and GLIMPSE data.

We modeled the extinction in the Galactic bar region more
accurately and with higher resolution through using VVV data and
more importantly, combined this with 2MASS and GLIMPSE data. In
the first case, we complemented VVV observations for brighter
sources and, in the second used MIR colors that allow a more
precise determination of extinction. The extinction model used
should not significantly change our bar representation (Figs. 6
and 7). The impact of any other extinction model should only
appear on a small scale as an effect of the resolution not
modifying the large structure of long bar as presented in this
work. Even if other extinction laws were considered the
normalization of the counts would be affected, but the shape would
be similar.

In terms of the parameters of the Galactic bar, our results
support those of previous studies (H00, LC01, LC07 and references
therein). But the spatial description of the long Galactic bar as
presented in this work (Figs. 6 and 7 lower panels) surpasses what
has been known previously, and allows us to better identify its
structure, not only in resolution but also in completeness.
Nevertheless, previous works by other authors have allowed us to
clearly identify structures along the bar, and this has allowed us
to identify and compare structure and regions. For instance, in
the region from $\ell \sim -14.0^\circ$ to $-12.0^\circ$ that bar
was possibly first detected reported by LC01. However, the limits
shown in this work are now much more clearly defined than in this
or any other previous study, using data from the DENIS, CAIN and
2MASS surveys. The same goes for all negative bar regions.

In the same way, with resulting longitudinal profiles made with
resolution (keeping robustness) at 1 degree interval, instead of 5
degrees as in previous works in the NIR, we more clearly reveal a
structure extending to $\ell \sim -14^\circ$. Our work can also be
used to constrain parameters by using models that separate bulge
and bar counts. The latitudinal profiles allow us to obtain the
centroid (as well as the bar`s vertical thickness) variation of
the stellar distribution with longitude in the $K_{\rm s}$ band
that shows a clear feature with a minimum at $\ell \approx
-13.8^\circ$.

\begin{acknowledgements}
We thank both the referee Dr Chris Flynn and the anonymous referee
for useful, valuable, and detailed comments on the manuscript that
also improved the paper`s clarity. We gratefully acknowledge use
of data from the ESO Public Survey program ID 179.B-2002 taken
with the VISTA telescope, data products from the Cambridge
Astronomical Survey Unit, and funding from the FONDAP Center for
Astrophysics 15010003, the BASAL CATA Center for Astrophysics and
Associated Technologies PFB-06, the MILENIO Milky Way Millennium
Nucleus from the Ministry of Economy's ICM grant P07-021-F, and
the FONDECYT from CONICYT. Eduardo Am\^ores obtained financial
support for this work from the Funda\c{c}\~{a}o para a Ci\^{e}ncia
e Tecnologia (FCT) under the grants SFRH/BPD/42239/2007 and
PTDC/CTE-SPA/118692/2010 and also CNPq (311838/2011-1). He also
acknowledges the hospitality and courtesy during his visit to the
IAC, as well as for the partial support for that. EBA thanks the
SIM-GRIDPT computing center funded under project
REDE/1522/RNG/2007. This work also made use of the computing
facilities of the Laboratory of Astroinformatics (IAG/USP,
NAT/Unicsul), where the cost was covered by the Brazilian agency
FAPESP (grant 2009/54006-4) and the INCT-A. MLC was supported by
the grant AYA2007-67625-CO2-01 of the Spanish Science Ministry.
CGF gratefully acknowledges the funding from the Spanish
Ministerio de Ciencia e Innovaci\'on (MCINN) under
AYA2010-21697-C05-5 and the Consolider-Ingenio 2010 Program grant
CSD2006-00070: First Science with the GTC
(http://www.iac.es/consolider-ingenio-gtc). This publication made
use of data products from the Two Micron All Sky Survey, which is
a joint project of the University of Massachusetts and the
Infrared Processing and Analysis Center/California Institute of
Technology, funded by the National Aeronautics and Space
Administration and the National Science Foundation. This work is
based in part on observations made with the Spitzer Space
Telescope, which is operated by the Jet Propulsion Laboratory,
California Institute of Technology under a contract with NASA.
\end{acknowledgements}

\clearpage
\newpage

\appendix

\section{Testing the method of extinction correction for the star counts}

In \S 3.1, we have proposed a simple algorithm to correct the
$K_{\rm s}-$band star counts from extinction, in which we only
need the observed magnitudes in $J$ and $K_{\rm s}$. There is some
discussion of the method in the literature (e.g., Alard 2001,
L\'opez-Corredoira et al. 2001) or a similar method for redder
wavelengths (e.g., M11), and the logic is clear: the
\textit{average} reddening is proportional to the extinction, thus
the average $(J-K_s)$ gives us information about the average
extinction along the line of sight. In our case, looking at the
central regions ($|\ell |\le 20^\circ $, $|b|<2.5^\circ $) and
excluding the sources with $(J-K_s)<0.5$, the contribution of disk
sources is low. So, star counts are dominated by sources from the
long bar or bulge+long bar. Since these structures have low
dispersion of distances, the dispersion of extinction of the
sources will not be high; as a result the application of the
method of correction will recover more or less the average counts
up to a limiting magnitude. In this section, we carry out some
simulations to show that the method of correction approximately
recovers the equivalent counts if we have no extinction.

For our simulation, we take a model of the Galaxy with a stellar
density of a disk from L\'opez-Corredoira et al. (2004), a bulge
from L\'opez-Corredoira et al. (2005, model 2), and a long bar
from LC07. Spiral arms and halo contributions are negligible so we
do not include them here. In Fig. A.1., we plot the face-on image
of the Galaxy according to this model. It is not important here
whether this is a true representation of the Galaxy or not. At
present, it is just a density distribution to test our method of
extinction correction.

We use a model of populations from Wainscoat et al. (1992) to
characterize the distribution of magnitudes and colors in each
point of the space. We take the disk populations, since the
differences with the older population of the bulge only affects
very bright giants or supergiants (L\'opez-Corredoira et al.
2005), which are brighter than our range of magnitudes. The long
bar is supposed to be an intermediate population, so again a
distribution of populations like the disk is appropriate.
Nonetheless, we insist that this is just a toy model for
evaluating the method of extinction correction, so variations in
this assumption are not important here. With the mentioned density
distribution and the population distribution, we integrate along
the line of sight (there is a factor $r^2$ in the integration),
and we get the synthetic color-magnitude diagram of Fig. A.2.
(left) at $\ell =-10^\circ $, $b=1^\circ $. We introduce a toy
model of extinction. We assume a distribution of dust as

\begin{equation}
\rho _{dust}=A_d\exp\left(-\frac{R}{h_R}-\frac{|z|}{h_z}\right)
,\end{equation}

\noindent where $h_R=3$ kpc, $h_z=0.1$ kpc, and extinction in
$K_{\rm s}$ in the solar neighborhood of 0.07 mag/kpc, which gives
2.8 magnitudes of extinction up to the center of the Galaxy at
distance 8.0 kpc. Again, whether this distribution of dust is
exact or not is not important for our exercise. In Fig. A.2.
(right) we plot the same color-magnitude diagram as before but
introducing this extinction, assuming
$\frac{A_{K_s}}{A_J-A_{K_s}}=\frac{3}{5}$ (see 5th paragraph of
Section 3.1).

Now, we calculate the star counts with $(J-K_s)\ge 0.5$ without
extinction in Fig. A.2. (left) and with extinction (Fig. A.2.
right) applying the correction of Equation (A.1) with
$(J-K_s)_0=1.0$. The results are plotted in Fig. A.3., for three
different lines of sight with three different total extinctions up
to the semiaxis of the long bar: respectively 2.78 mag up to a
distance of 8 kpc for $\ell =0$, $b=0$; 1.24 mag up to at distance
9.7 kpc for $\ell =-10^\circ $, $b=1^\circ $; 0.47 mag up to a
distance 12.9 kpc for $\ell =-20^\circ $, $b=2^\circ $. We
therefore test different longitudes, latitudes, and extinctions.

As we can see in Fig. A.3., the method approximately recovers the
counts that would be produced without extinction, within small
errors typically of 5-10\%, which are not important for our
purposes. The variation of
$\frac{A_{K_s}}{A_J-A_{K_s}}=\frac{3}{5}$ will also introduce some
extra error.

\begin{figure}
\vspace{1cm} {\par\centering
\resizebox*{7.5cm}{7.5cm}{\includegraphics{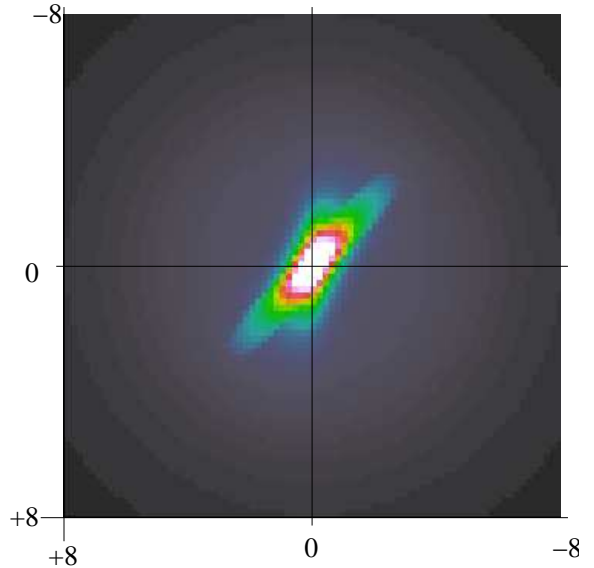}}
\par\centering}
\caption{Face-on view (in units of kpc) of the Model of the Galaxy
(see text in Appendix A) used to test the extinction correction
method.} \label{Fig:model}
\end{figure}

\bigskip
\clearpage

\begin{figure*}
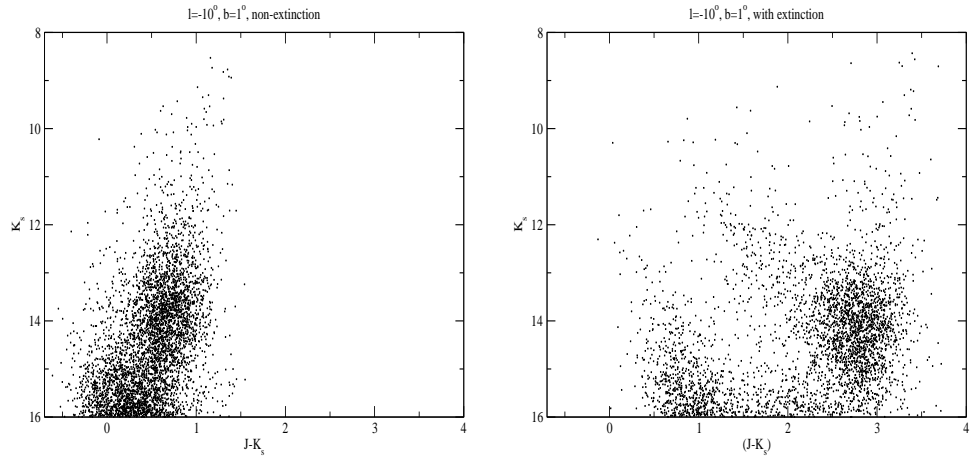

\vspace{1cm} {\par\centering
\resizebox*{6cm}{6cm}{\includegraphics{synthCMa.eps}}\hspace{.5cm}
\resizebox*{6cm}{6cm}{\includegraphics{synthCMb.eps}}
\par\centering}
\caption{Synthetic color-magnitude diagram in the direction $\ell
=-10^\circ $, $b=1^\circ $ (left) without extinction; (right) with
extinction.} \label{Fig:synthCM}
\end{figure*}

\bigskip
\clearpage

\begin{figure}
\vspace{1cm} {\par\centering
\resizebox*{7.5cm}{7.5cm}{\includegraphics{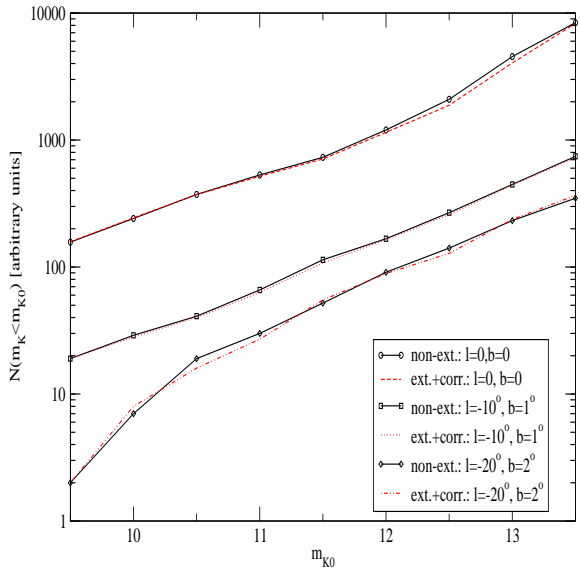}}
\par\centering}
\caption{Calculations of the counts (arbitrary normalization) for
the model in different regions without extinction, and with
extinction+correction of the extinction.} \label{Fig:testcorrext}
\end{figure}

\end{document}